\documentclass[iop,twocolappendix,numberedappendix,appendixfloats, tighten]{emulateapj}

\usepackage{graphicx}
\usepackage[space]{grffile}
\usepackage{latexsym}
\usepackage{amsfonts,amsmath,amssymb}
\usepackage{url}
\usepackage[utf8]{inputenc}
\usepackage{fancyref}
\usepackage{textcomp}
\usepackage{longtable}
\usepackage{multirow,booktabs}
\usepackage{subfigure}
\usepackage{natbib}
\usepackage{color}

\usepackage[backref,breaklinks,colorlinks,citecolor=blue]{hyperref}

\newcommand{\msol}{M$_\odot$}

\newcommand{\logmass}{$\mathrm{log_{10}(M_*/M_\odot)}$}
\newcommand{\zsol}{Z$_\odot$}

\newcommand{\OVI}{[\hbox{{\rm O}\kern 0.1em{\sc vi}}]}
\newcommand{\Lalpha}{Ly-$\alpha$}
\newcommand{\NV}{\hbox{{\rm N}\kern 0.1em{\sc v}}}
\newcommand{\SiIV}{\hbox{{\rm Si}\kern 0.1em{\sc iv}}}
\newcommand{\OIV}{[\hbox{{\rm O}\kern 0.1em{\sc iv}}]}
\newcommand{\NIV}{[\hbox{{\rm N}\kern 0.1em{\sc iv}}]}
\newcommand{\CIV}{\hbox{{\rm C}\kern 0.1em{\sc iv}}}
\newcommand{\HeII}{\hbox{{\rm He}\kern 0.1em{\sc ii}\kern 0.1em{$\lambda1640$} }}
\newcommand{\OIII}{[\hbox{{\rm O}\kern 0.1em{\sc iii}}]{$\lambda5007$}}
\newcommand{\NIII}{[\hbox{{\rm N}\kern 0.1em{\sc iii}}]}
\newcommand{\AlIII}{\hbox{{\rm Al}\kern 0.1em{\sc iii}}}
\newcommand{\SiIII}{\hbox{{\rm Si}\kern 0.1em{\sc iii}}}
\newcommand{\CIII}{\hbox{{\rm C}\kern 0.1em{\sc iii}]}}
\newcommand{\NeIV}{[\hbox{{\rm Ne}\kern 0.1em{\sc iv}}]}
\newcommand{\MgII}{\hbox{{\rm Mg}\kern 0.1em{\sc ii}}}

\newcommand{\CII}{[\hbox{{\rm C}\kern 0.1em{\sc ii}]}}

\newcommand{\He}{\hbox{{\rm He}\kern 0.1em{\sc ii}\kern 0.1em{$\lambda1640\lambda4686$}}}

\newcommand{\Halpha}{H$\alpha$}
\newcommand{\Hbeta}{H$\beta$}
\newcommand{\SII}{[\hbox{{\rm S}\kern 0.1em{\sc ii}}]$\lambda6717\lambda6731$}
\newcommand{\NII}{[\hbox{{\rm N}\kern 0.1em{\sc ii}}]}
\newcommand{\OII}{[\hbox{{\rm O}\kern 0.1em{\sc ii}}]}

\newcommand{\MgI}{\hbox{{\rm Mg}\kern 0.1em{\sc i}}}
\newcommand{\FeII}{\hbox{{\rm Fe}\kern 0.1em{\sc ii}}}
\newcommand{\HII}{{\ion{H}{2}}}

\newcommand{\OI}{\hbox{{\rm O}\kern 0.1em{\sc i}}}
\newcommand{\NeII}{[\hbox{{\rm Ne}\kern 0.1em{\sc ii}}] }
\newcommand{\NaI}{[\hbox{{\rm Na}\kern 0.1em{\sc i}}] }
\newcommand{\NeIII}{[\hbox{{\rm Ne}\kern 0.1em{\sc iii}}] }

\newcommand{\Av}{$A(V)$}

\newcommand{\NMAD}{$\sigma_{\mathrm{NMAD}}$}

\newcommand{\xiion}{$\xi_{ion}$}

\newcommand{\boxfil}{$\mathrm{[340]-[550]}$}

\slugcomment{To appear in The Astrophysical Journal (ApJ)}

\shorttitle{ZFIRE $\xi_{ion}$ at $z\sim2$}
\shortauthors{Themiya Nanayakkara} 

\begin{document}



\title{Reconstructing the observed ionizing photon production efficiency at $z\sim2$ using stellar population models}


\author{Themiya Nanayakkara\altaffilmark{1,*}}
\author{Jarle Brinchmann\altaffilmark{2}}
\author{Karl Glazebrook\altaffilmark{3}}
\author{Rychard Bouwens\altaffilmark{1}}
\author{Lisa Kewley\altaffilmark{4,5}}
\author{Kim-Vy Tran\altaffilmark{5,6,7,8}}
\author{Michael Cowley\altaffilmark{9,10}}
\author{Deanne Fisher\altaffilmark{3}}
\author{Glenn G. Kacprzak\altaffilmark{3}}
\author{Ivo Labbe\altaffilmark{3}}
\author{Caroline Straatman\altaffilmark{11}}

\altaffiltext{1}{Leiden Observatory, Leiden University, PO Box 9513, 2300 RA Leiden, Netherlands.}
\altaffiltext{*}{themiyananayakkara@gmail.com}
\altaffiltext{2}{ Instituto de Astrof{\'\i}sica e Ci{\^e}ncias do Espaço, Universidade do Porto, CAUP, Rua das Estrelas, PT4150-762 Porto, Portugal.}
\altaffiltext{3}{Centre for Astrophysics and Supercomputing, Swinburne University of Technology, Hawthorn, Victoria 3122, Australia.}
\altaffiltext{4}{Research School of Astronomy and Astrophysics, The Australian National University, Cotter Road, Weston Creek, ACT 2611, Australia.}
\altaffiltext{5}{ARC Centre of Excellence for All Sky Astrophysics in 3 Dimensions (ASTRO 3D), Canberra, Australian Capital Territory 2611, Australia.}
\altaffiltext{6}{School of Physics, University of New South Wales, Sydney, NSW 2052, Australia.}
\altaffiltext{7}{Department of Physics and Astronomy, Texas A\&M University, College Station, TX, 77843-4242 USA.}
\altaffiltext{8}{George P. and Cynthia Woods Mitchell Institute for Fundamental Physics and Astronomy, Texas A\&M University, College Station, TX, 77843-4242.}
\altaffiltext{9}{Centre for Astrophysics, University of Southern Queensland, West Street, Toowoomba, QLD 4350, Australia.}
\altaffiltext{10}{School of Chemistry, Physics and Mechanical Engineering, Queensland University of Technology, Brisbane, QLD 4001, Australia.}
\altaffiltext{11}{Sterrenkundig Observatorium, Universiteit Gent, Krijgslaan 281 S9, B-9000 Gent, Belgium.}

\begin{abstract}
The ionizing photon production efficiency, $\xi_{ion}$, is a critical parameter that provides a number of physical constraints to the nature of the early Universe, including the contribution of galaxies to the timely completion of the reionization of the Universe.   
Here we use KECK/MOSFIRE and ZFOURGE multi-band photometric data to explore the $\xi_{ion}$ of a population of galaxies at $z\sim2$ with $log_{10}(M_*/M_\odot)\sim9.0-11.5$. 
Our 130 \Halpha\ detections show a median $log_{10}(\xi_{ion}[Hz/erg])$ of $24.8\pm0.5$ when dust corrected using a  Calzetti et al. (2000) dust prescription.  
Our values are typical of mass/magnitude selected $\xi_{ion}$ values observed in the $z\sim2$ Universe.
Using BPASSv2.2.1 and Starburst99 stellar population models with simple parametric star-formation-histories (SFH), we find that even with models that account for effects of stellar evolution with binaries/stellar rotation, model galaxies at $log_{10}(\xi_{ion}[Hz/erg])\lesssim25.0$ have low H$\alpha$ equivalent widths (EW) and redder colors compared to our $z\sim2$ observed sample. 
We find that introducing star-bursts to the SFHs resolve the tension with the models, however, due to the rapid time evolution of $\xi_{ion}$, H$\alpha$ EWs, and rest-frame optical colors, our Monte Carlo simulations of star-bursts show that random distribution of star-bursts in evolutionary time of galaxies are unlikely to explain the observed distribution. 
Thus, either our observed sample is specially selected based on their past SFH or stellar models require additional mechanisms to reproduce the observed high UV luminosity of galaxies for a given production rate of hydrogen ionizing photons. 
\end{abstract}


\keywords{galaxies: evolution -- galaxies: fundamental parameters -- galaxies: high-redshift -- galaxies: ISM -- galaxies: star formation}

\section{Introduction}
\label{sec:introduction}

Current observational constraints suggest that the reionization of the Universe occurred between $z\sim20-6$  through the escape of ionizing photons (Lyman continuum leakage) from young stellar populations in galaxies \citep{Bouwens2015,Finkelstein2015b,Robertson2015}. 
However, the exact source of these photons that are predominantly responsible for reionization is still under debate.
To put constraints on mechanisms that drove the reionization, it is important to understand properties of the massive stars in this era and link how the production of ionizing photons from these stars influenced the ionization of surrounding regions leading to cosmic reionization \citep[e.g.,][]{Barkana2006,Shin2008}.

The ionizing photon production efficiency, \xiion, is defined as the production rate of Lyman-continuum photons ($\mathrm{\lambda_{photon}<912\ \AA}$) per unit Ultra-Violet (UV) continuum luminosity measured at 1500 \AA. 
\xiion\ provides a measure of hydrogen ionizing to non-ionizing photon production rates and therefore is a measure of the ratio of massive to less massive stars in stellar populations.
\xiion\ combined with the UV luminosity density and the escape fraction of ionizing photons is required to compute the ionizing emissivity from galaxies to determine if and how galaxies drove the reionization of the Universe \citep[e.g.,][]{Kuhlen2012,Naidu2019}. 
Additionally, \xiion\  is an ideal measurable to compare with stellar population model predictions, especially at the peak of the cosmic star-formation rate density ($z\gtrsim2$).

A direct measure of \xiion\ requires a flux measurement to be obtained below the Lyman limit.  
Even at $z\sim2$ this requires extremely deep observations, suffers from high systematic errors, and given the high IGM absorption from observed sight-lines, can only be done on stacked samples (for a thorough analysis see \citealt{Steidel2018}, also see \citealt{Reddy2016b}).
Additionally, stellar population synthesis models can be used to calibrate the the rest-UV continuum slope, $\beta$, with \xiion, and multiple studies have used the observed $\beta$ to infer \xiion\ of high-$z$ galaxies \citep[,][]{Robertson2013,Bouwens2015b}. 
However, $\beta$ is also sensitive to dust, metallicity, and star-formation histories \citep[SFHs,  e.g.,][]{Reddy2018} of galaxies and thus inferred \xiion\ values are influenced by related uncertainties. 
\xiion\ measurements from UV metal lines \citep{Stark2017} requires deep exposures and suffer from further stellar population and photo-ionization uncertainties.

In ionization bounded \HII\ regions, under dust free Case B recombination, \Halpha\ emission is directly proportional to the number of Lyman continuum photons produced by hot young stars and has been used to estimate \xiion\ \citep[e.g.,][]{Bouwens2016c,Matthee2017b,Nakajima2017,Shivaei2018}. 
Studies that use narrow band imaging suffer from strong line contamination due to the close proximity of \NII\ (and \SII\ in the case of broad band imaging), which is usually corrected for using either empirical or model calibrations \citep{Bouwens2016c,Matthee2017b}. However, such calibrations are not well tested at $z\gtrsim2$ and due to harder ionizing fields and variations in element abundances, emission line ratios have shown to evolve from local calibrations \citep[e.g.,][]{Steidel2014,Kewley2016,Strom2017}. This could introduce systematic biases to line flux estimates. 
Additionally, accurate dust corrections to UV and nebular \Halpha\ flux require a combination of multi-wavelength photometry and Balmer line ratios (see \citealt{Shivaei2018}), thus spectroscopic measurements are crucial to obtain accurate estimates of the number of ionizing photons.

In this analysis, we take advantage of the recombination nature of the nebular \Halpha\ emission line to estimate the amount of ionizing photons produced within galaxies. 
We combine MOSFIRE \citep{McLean2012} spectroscopic observations by the ZFIRE survey \citep{Tran2015,Nanayakkara2016} with multi-wavelength photometry by the ZFOURGE survey \citep{Straatman2016} to compute the ionizing photon production efficiency of a population of galaxies at $z\sim2$. The paper is structured as follows: in Section \ref{sec:data} we present our sample, in Section \ref{sec:analysis} we present an analysis of the \xiion\ measurements with observed/derived properties of our sample, in Section \ref{sec:discussion} we briefly discuss our results, and present our conclusions in  Section \ref{sec:conclusion}.
Unless otherwise stated, we assume a \citet{Chabrier2003} IMF and a cosmology with  H$_{0}= 70$ km/s/Mpc, $\Omega_\Lambda=0.7$ and $\Omega_m= 0.3$. All magnitudes are expressed using the AB system \citep{Oke1983}.


\section{Sample selection and Results}
\label{sec:data}

\subsection{Survey description}
\label{sec:survey}

The spectroscopic data used in this analysis was obtained as a part of the ZFIRE survey (PIs K. Glazebrook, L. Kewley, K. Tran) which utilized the MOSFIRE instrument on Keck-I telescope to obtain rest-frame optical spectra of mass/magnitude selected samples of galaxies around galaxy rich environments at $z=1.5-2.5$ \citep{Yuan2014,Kacprzak2015,Tran2015,Alcorn2016}. 
A thorough description of survey goals, sample selection, data reduction, flux calibration, and line flux measurements is presented in \citet{Nanayakkara2016}. 
The ZFIRE sample in the COSMOS field \citep{Scoville2007} comprises of all 134 galaxies observed in MOSFIRE $K$ band with secure \Halpha\ detections ({\tt conf=3}, redshift determined by multiple emission lines) between $1.90<z<2.67$ with a $5\sigma$ line flux detection level $\sim3\times10^{-18}$ erg/s/cm$^2$/\AA. 
The 80\% stellar mass and $Ks$ completeness of this sample are respectively, $log_{10}(M_*/M_\odot)>9.3$ and $Ks<24.11$. 
We remove 4 galaxies flagged as AGN by \citet{Cowley2016}. These AGN selections are based on based on infrared color-color classifications of \citep{Messias2012}, \citet{Rees2016} radio AGN activity index, and X-ray AGN selection criteria of \citep{Szokoly2004} and we remove these galaxies from our sample. 
We consider the remaining 130 galaxies as our primary sample.
\citet{Nanayakkara2016} showed that the \Halpha\ selected sample contains no significant systematic biases towards SFH, stellar mass, and Ks band magnitude based on the parent ZFOURGE sample \citep{Straatman2016}.

ZFIRE spectroscopic data supplements the ZFOURGE survey (PI I. Labbe),  a $Ks$-selected deep 45 night photometric legacy survey carried out using the purpose built {\tt FourStar} imager \citep{Persson2013} in the 6.5-m Magellan Telescope. The survey covers 121 arcmin$^2$ in each of the COSMOS, UDS \citep{Beckwith2006}, and CDFS \citep{Giacconi2001} legacy fields reaching a 5$\sigma$ depth of $Ks\leq25.3$ AB and is complemented by the wealth of public multi-wavelength photometric data (UV to far infrared) available in these fields \citep{Straatman2016}.


\subsection{\xiion\ computation and dust corrections}
\label{sec:xiion_computation}

For our analysis we select all galaxies from the ZFIRE survey in the COSMOS field between $1.90<z<2.67$ with a {\tt conf=3}  and a \Halpha\ signal to noise (S/N) $>5$.
We define \xiion\ as:
\begin{equation}
\label{eq:xiion_def}
\xi_{ion} = \frac{N(H)}{L_{UV}}\ [Hz/erg]
\end{equation} 
where $N(H)$ is the production rate of H ionizing photons per $s$, and $L_{UV}$ is the intrinsic UV continuum luminosity at 1500 \AA. 

In order to obtain the observed $L_{UV}$, we first refit ZFOURGE photometry using FAST++ \citep{Schreiber2018} at the spectroscopic redshifts and compute the UV luminosity at rest-frame  1500 \AA\ by fitting a power law function to the best fit spectral energy distribution (SED) model between $\Delta \lambda=1400-1600$ \AA. We use the exponential of the same power law as the UV continuum slope $\beta$.  

Best fit \Av\ values and stellar masses from FAST++ are computed using \citet{Bruzual2003} stellar population models with a \citet{Chabrier2003} IMF, a truncated SFH with a constant and an exponentially declining SFH component, and a \citet{Calzetti2000} dust law. 
Galaxies are fixed at the spectroscopic redshift similar to \citet{Nanayakkara2016}, however, we allow the stellar metallicity to vary as a free parameter between $Z=0.004-0.02$, within which readily computed SFH models are available in FAST++.  
We use the FAST++ computed \Av\ values to obtain the intrinsic UV luminosity using the \citet{Calzetti2000} dust law. 
Additionally, in Table \ref{tab:SED_fitting_params} we show that the choice of the SFH and the dust attenuation law in FAST++ may contribute up to $\sim0.1\pm0.3$ and $\sim0.02\pm0.1$ systematic offset to $\beta$ and UV magnitude measurements, respectively.

\begin{deluxetable*}{llccrr}
\tabletypesize{\scriptsize}
\tablecaption{ Role of the assumed SFH and dust law in FAST++}  
\tablecomments{ Here we show the median offset and \NMAD\ of $\beta$ and M(UV) between different SFHs and dust laws computed using FAST++. Throughout the analysis best-fit SEDs derived assuming a truncated SFH with a \citet{Calzetti2000} dust law is used.  
\label{tab:SED_fitting_params}}
\tablecolumns{6}
\tablewidth{0pt} 
\startdata
\hline \hline \\ 
SFH  1				& SFH 2     					 &  Dust Law 1   				&  Dust Law 2 				&   $\Delta\beta$		& $\Delta$M(UV)    	            	\\
\hline \\ 
Truncated 			& Exponentially declining 		&    \citet{Calzetti2000}  		&  \citet{Calzetti2000}     &	$-0.003\pm 0.10$  	& $0.0007\pm 0.06$  				\\
Truncated 			& Delayed $\tau$ 				&    \citet{Calzetti2000}  		&  \citet{Calzetti2000}     &	$ 0.015\pm 0.11$  	& $0.007\pm 0.06$  					\\
Truncated 			& Truncated 					&    \citet{Calzetti2000}  		&  \citet{Cardelli1989}     &	$-0.05\pm 0.30$  	& $0.008\pm 0.10$  					\\
Truncated 			& Truncated 					&    \citet{Calzetti2000}  		&  \citet{Kriek2013}     	&	$-0.07\pm 0.12$  	& $-0.02\pm 0.05$  					\\
\end{deluxetable*}

$N(H)$ is computed following dust free Case B recombination at an electron density of $n_{e}=10^3\ cm^{-3}$ and temperature of $T=10,000\ K$ assuming no escape of ionizing photons:
\begin{equation}
\label{eq:N_H}
N(H) = \frac{L(H\alpha)}{C_B} [s^{-1}]
\end{equation}
where $C_B ={1.36}\times 10^{-12} erg = (\alpha_{effH\alpha}/\alpha_{effCB}) \times h\nu_{H\alpha}$ with $\alpha_{effH\alpha}=1.17\times10^{-13}\ cm^3/s$, $\alpha_{effCB}=2.59\times10^{-13}\ cm^3/s$, and $h\nu_{H\alpha}=3.03\times10^{-12}\ erg$ \citep{Draine2011}. 
The intrinsic \Halpha\ luminosity ($L(H\alpha)\ [erg/s]$) is computed using the dust corrected observed \Halpha\ flux from the ZFIRE spectra as described below.

\begin{figure*}
\includegraphics[scale=0.8]{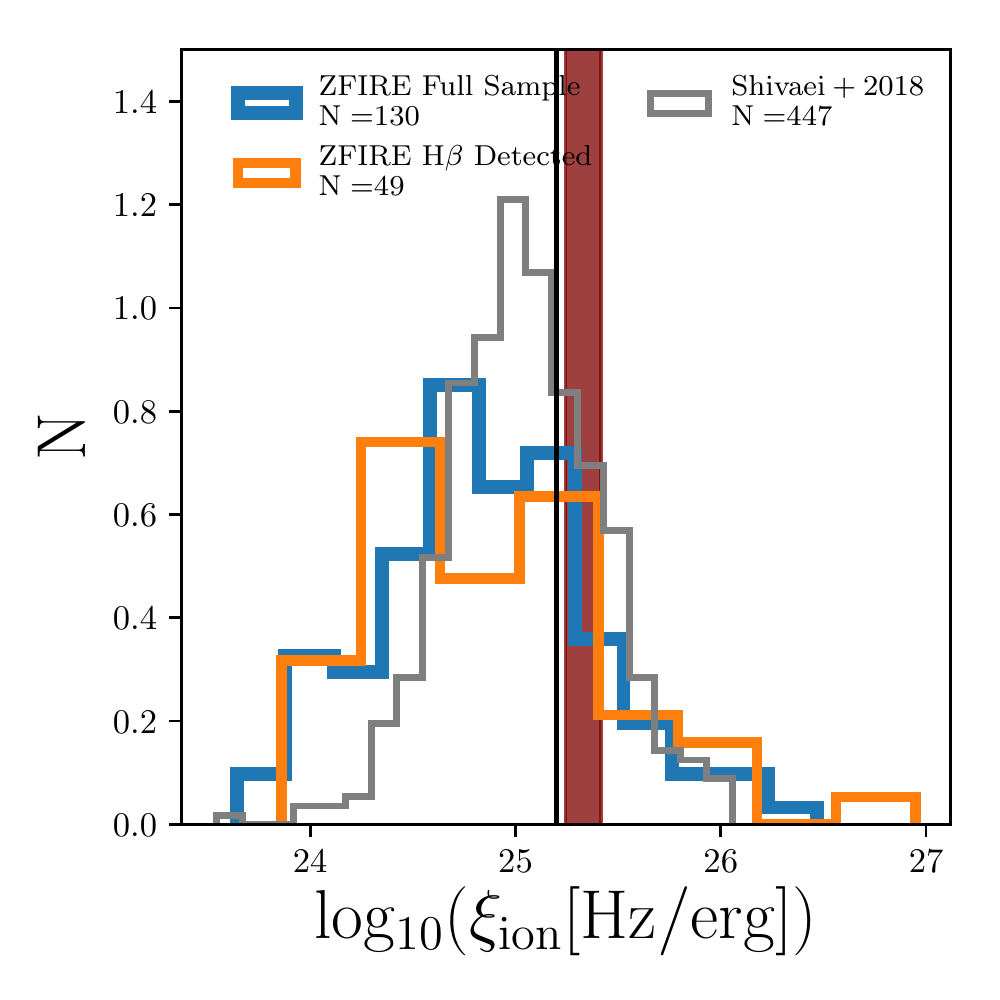}
\includegraphics[scale=0.8]{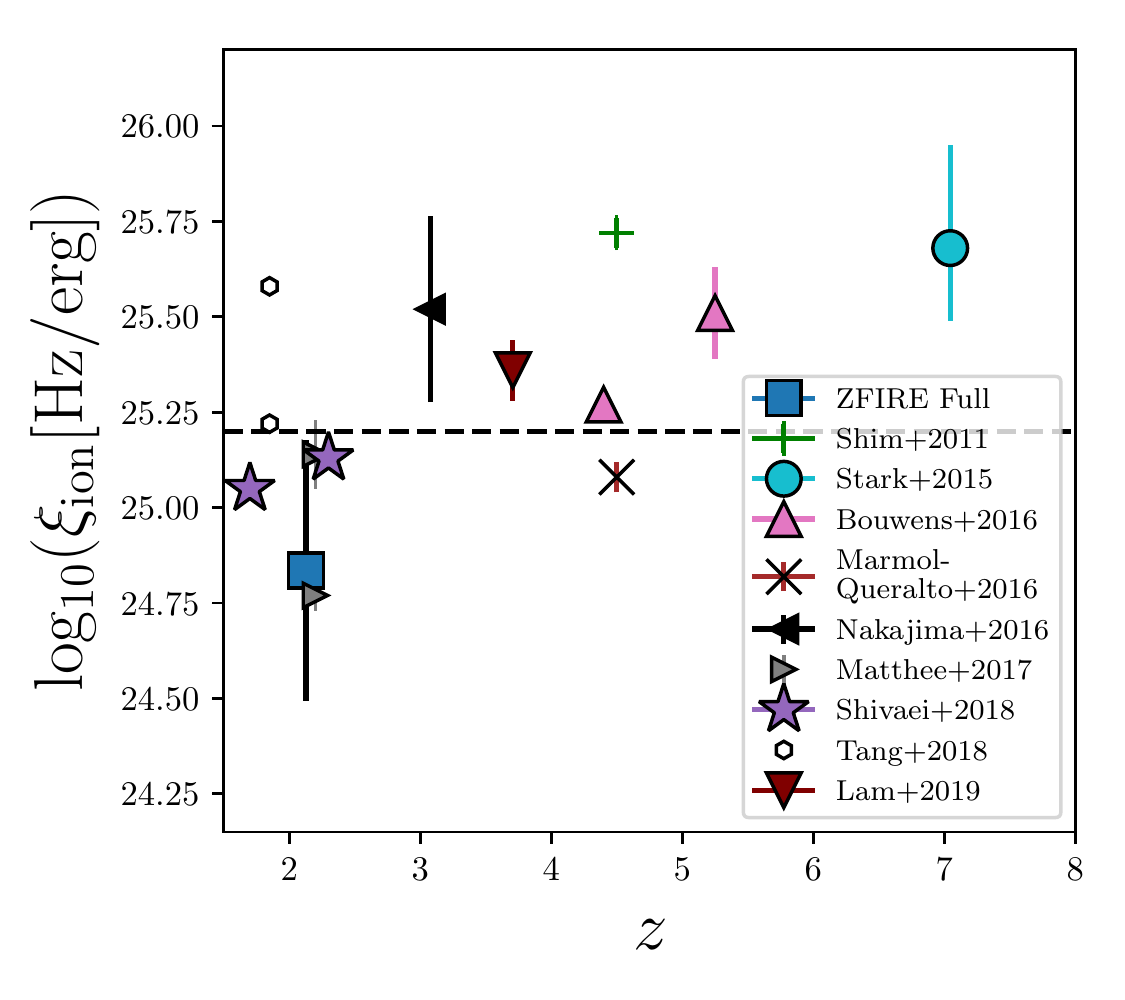}
\caption{ {\bf Left:} The \xiion\ distribution of our sample. We show histograms of set A and set B galaxies along with the \citet{Shivaei2018} $z\sim2$ sample. 
The maroon shading shows the \xiion\ distribution for a BPASSv2.2.1 binary stellar population model with a  \citet{Salpeter1955} like IMF with an upper mass cut at 300\msol\ at solar metallicity between 10-100 Myr of age. 
As a reference we show the  \citet{Robertson2013} $log_{10}(\xi_{ion}[Hz/erg])=25.2$ (the \xiion\ needed to reionize the Universe by $z\sim6$) as a black vertical line.
{\bf Right:} The distribution of \xiion\ as a function of $z$ for galaxies in our sample and a selected subset of galaxies from literature: \citet{Shim2011,Stark2015c,Bouwens2016c,Marmol-Queralto2016,Nakajima2016,Matthee2017b,Shivaei2018,Tang2018,Lam2019a} are shown for comparison. The dashed black horizontal line is the  \citet{Robertson2013} $log_{10}(\xi_{ion}[Hz/erg])=25.2$ value. 
\label{fig:xiion_distribution_z}}
\end{figure*}


\subsubsection{Deriving nebular emission line corrections}
\label{sec:sample_A_B}

We divide the sample into two sets and apply dust corrections following \citet{Cardelli1989} and  \citet{Calzetti2000} dust laws for the nebular and stellar regions, respectively, as commonly done in literature at $z\sim2$ \citep[e.g.,][]{Shivaei2018}. 
We define our full sample as set A, which contains 130 galaxies with MOSFIRE \Halpha\ detections observed in the K band.
Only a subset of these galaxies were observed in MOSFIRE H band to obtain \Hbeta\ detections. 
We define the sub-sample of 49 galaxies from set A with \Hbeta\ detections (S/N $>=3$), as set B.

For our set B galaxies, we use the \citet{Cardelli1989} attenuation curve with the Balmer decrement values of the individual galaxies to obtain intrinsic \Halpha\ luminosities following the  Case B value of $f(H\alpha)/f(H\beta)=2.86$.

We stack galaxies in set A with MOSFIRE H band observations in four \Halpha\ SFRs bins and compute an average balmer decrement for each bin. Then we compute the intrinsic \Halpha\ luminosity for galaxies in each bin similar to set B using the average balmer decrement in each bin. Unless otherwise stated explicitly, we use this \xiion\ value for set A galaxies throughout the analysis.  
A summary of our galaxy sets is provided in Table \ref{tab:sample_details}.



\begin{deluxetable*}{lcccll}
\tabletypesize{\scriptsize}
\tablecaption{ \xiion\ sample definitions}  
\tablecomments{ Here we summarize the two samples used in our analysis. 
\label{tab:sample_details}}
\tablecolumns{6}
\tablewidth{0pt} 
\startdata
\hline \hline \\ 
Set Name  		& N of      &  UV luminosity dust    		&  \Halpha\ luminosity dust &   Balmer decrement 		& Median    	            		\\
            	& galaxies  &  correction law   			&  correction law     		& 	from					& log$_{10}$(\xiion\ [Hz/erg])  	\\
\hline \\ 
Set A 			& 130 		&    \citet{Calzetti2000}  		&  \citet{Cardelli1989}     &	\Halpha\ SFR stacks 	& $24.83\pm 0.49$  					\\
Set A 			& 130 		&    \citet{Calzetti2000}  		&  \citet{Cardelli1989}     &	$\beta$ stacks 			& $24.77\pm 0.43$  					\\
Set A 			& 130 		&    \citet{Calzetti2000}  		&  \citet{Cardelli1989}     &	UV magnitude stacks 	& $24.73\pm 0.49$  					\\
Set A 			& 130 		&    \citet{Calzetti2000}  		&  \citet{Cardelli1989}     &	Stellar mass stacks		& $24.79\pm 0.44$  					\\
Set A 			& 130 		&    \citet{Calzetti2000}  		&  \citet{Cardelli1989}     &	\OIII/\Halpha\ stacks	& $24.76\pm 0.45$  					\\
Set A 			& 130 		&    \citet{Calzetti2000}  		&  \citet{Cardelli1989}     &	UV+IR SFR stacks 		& $24.68\pm 0.46$  					\\
Set B			& 49 		&    \citet{Calzetti2000}  		&  \citet{Cardelli1989}     &	Individual observations & $24.79\pm 0.58$   				\\ 
\end{deluxetable*}


\subsection{The observed distribution of \xiion}
\label{sec:xiion_observed_distribution}


In Figure \ref{fig:xiion_distribution_z} we show the distribution of \xiion\ in our sample. 
Galaxies in set A are shown with dust corrections applied with average balmer decrements from \Halpha\ SFR stacks. The median of the distributions between set A and B are consistent within the scatter of the distribution (Table \ref{tab:sample_details}).

For comparison, in Figure \ref{fig:xiion_distribution_z} we also show the \citet{Shivaei2018} $z\sim2$ galaxy sample which has a median log$_{10}$(\xiion\ [Hz/erg])=$25.0\pm0.4$ with a stellar mass completeness at \logmass$\sim9.5$. 
Our set A sample shows a similar distribution of \xiion\ to the \citet{Shivaei2018} sample albeit with a slight bias towards low \xiion. 
Compared to \citet{Robertson2013} \xiion\ constraints to reionize the Universe by $z\sim6$, $\sim80\%$ of our set A galaxies fall below this limit. 
Additionally, we show the distribution of \xiion\ for BPASSv2.2.1 \citep{Eldridge2017} binary star constant SFH models at solar metallicity between  $10-100$ Myr from the onset of the star-formation. 
This range shows the realistic distribution of \xiion\ from the onset of a star-formation up to the point where the UV luminosity is stabilized in a constant SFH scenario.   
A majority of our galaxies in set A and set B have lower \xiion\ compared to these model predictions.

Our \xiion\ measurements are lower compared to model predictions and such differences could be driven by differences in the stellar population/ISM properties \citep[e.g.,][]{Kewley2019}, calibration uncertainties, and/or the choice of the dust attenuation curve. 
In \citet{Nanayakkara2016} we showed that the relative calibration between ZFIRE spectra and ZFOURGE photometry agrees within $\lesssim10\%$. We also visually inspected all spectra and the best-fit SEDs to determine if calibration offsets could drive the enhancement of \xiion\ and found in general good agreement between flux levels of the spectra and the SEDs. 
Therefore, we rule out calibration effects to have a dominant effect on the derived low \xiion\  of our sample. 
Even though we removed galaxies that showed evidence for AGN activity based on x-ray, infrared, and radio observations \citep{Cowley2016}, it is possible that our sample is contaminated by sub-dominant AGN. 
If AGN  primarily contribute to an excess of UV flux, the low \xiion\ of our sample could be driven by effects of sub-dominant AGN.  However, AGN emission will also increase the \Halpha\ emission and thus is unlikely to be a contributor to lowering \xiion\ of our sample.

In Figure \ref{fig:xiion_distribution_z}, we show the median distribution of \xiion\ presented in this study along with other studies from literature at $z\sim2$ and higher. 
There are no statistically significant differences between our galaxies and \citet{Shivaei2018} \xiion\ analysis, and is independent of the type of stacks used to compute the average balmer decrement. 
Further, in \citet{Nanayakkara2016}, we demonstrated that there were no statistically significant biases between the ZFOURGE COSMOS input sample and our \Halpha\ detected sample (Set A) in terms of stellar mass or $Ks$ magnitude.
\citet{Tang2018} compute \xiion\ for a highly selective sample of galaxies between $z\sim1.5-3$, which are selected  based on their extreme \OIII\ line properties and have stellar masses up to $\sim2$ dex lower than our stellar mass completeness.
A majority of galaxies probed by \citet{Nakajima2016} is of high mass (\logmass$>9.0$), however, the sample is based on \Lalpha\ emitters with extreme \OIII\ emission.
Therefore, it is not surprising that both \citet{Tang2018} and \citet{Nakajima2016} show on average higher \xiion\ compared to our sample.

Our \xiion\ are systematically lower than \xiion\ computed for $z\sim4$ galaxies by \citet{Lam2019a}, $z\sim5$ galaxies by \citet{Bouwens2016c}, and $z\sim7$ \Lalpha\ emitter by \citet{Stark2015c}. 
Thus, it is evident that our sample show \emph{typical} \xiion\ observed at $z\sim2$ and are lower than what is observed in galaxies at $z\gtrsim4$. This difference could be driven by a redshift evolution of \xiion \citep{Matthee2017b}, biases (i.e. mass incomplete samples) in sample selection of $z>4$ observations, and/or differences in SFHs. We discuss this further in Section \ref{sec:discussion_z_evolution}.

Next, we analyze the distribution of our sample with commonly probed correlations of \xiion\ in low and high-$z$ galaxies to investigate if such correlations also hold for our sample.


\section{Analysis}
\label{sec:analysis}

\subsection{Observed correlations of \xiion}
\label{sec:xi_ion_observed_correlations}

\begin{figure*}
\includegraphics[scale=0.6]{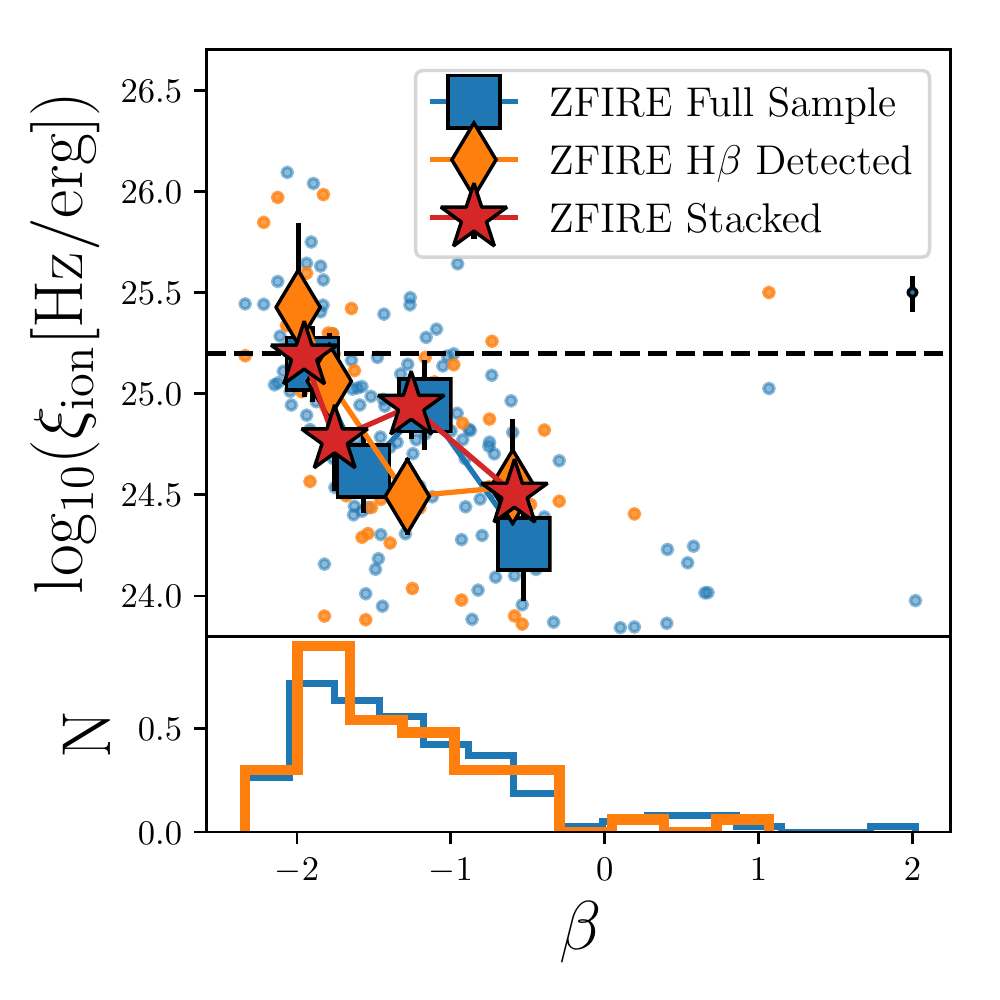}
\includegraphics[scale=0.6]{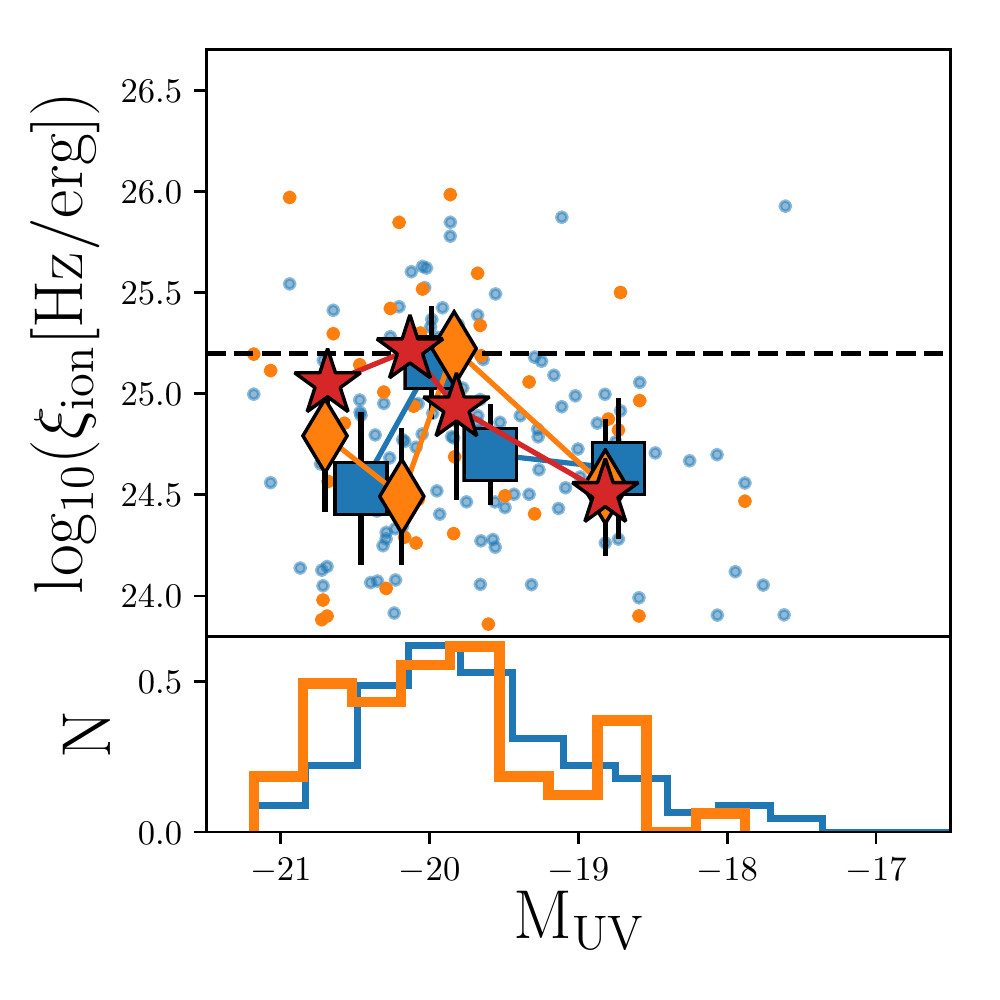}
\includegraphics[scale=0.6]{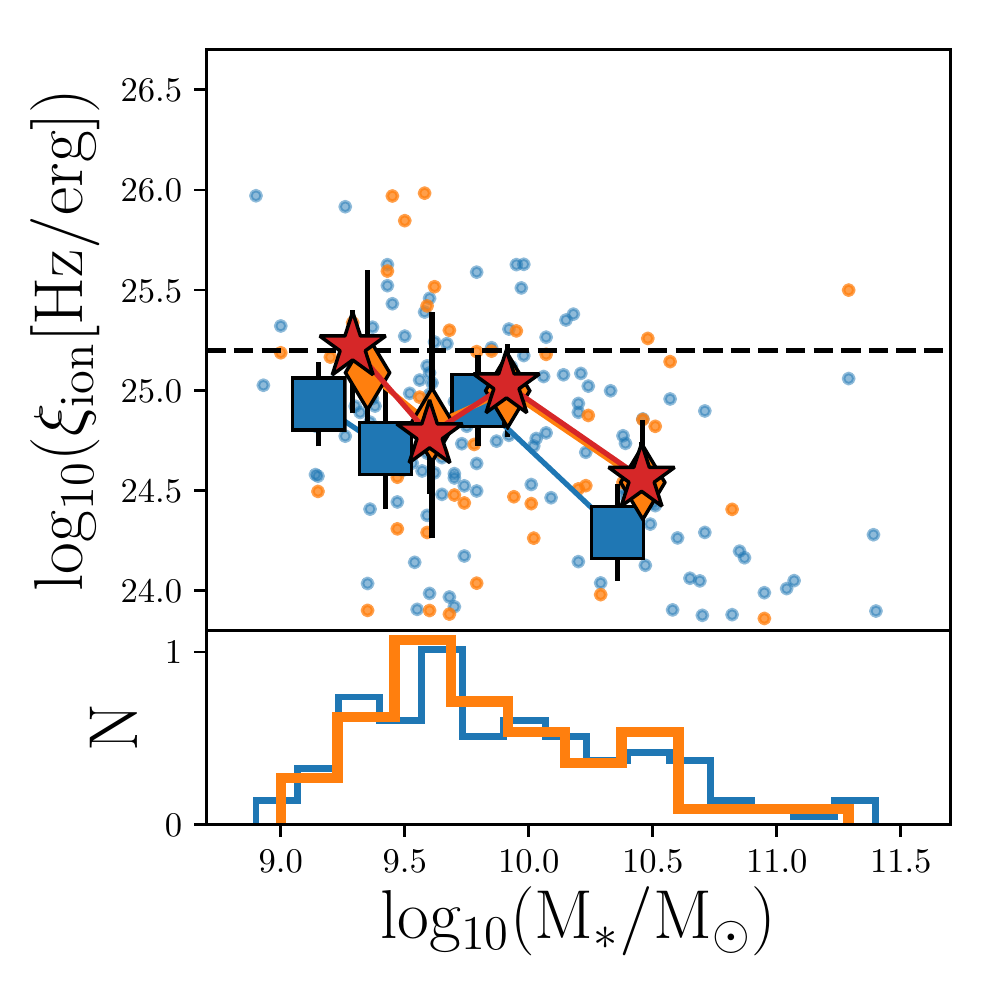}
\includegraphics[scale=0.6]{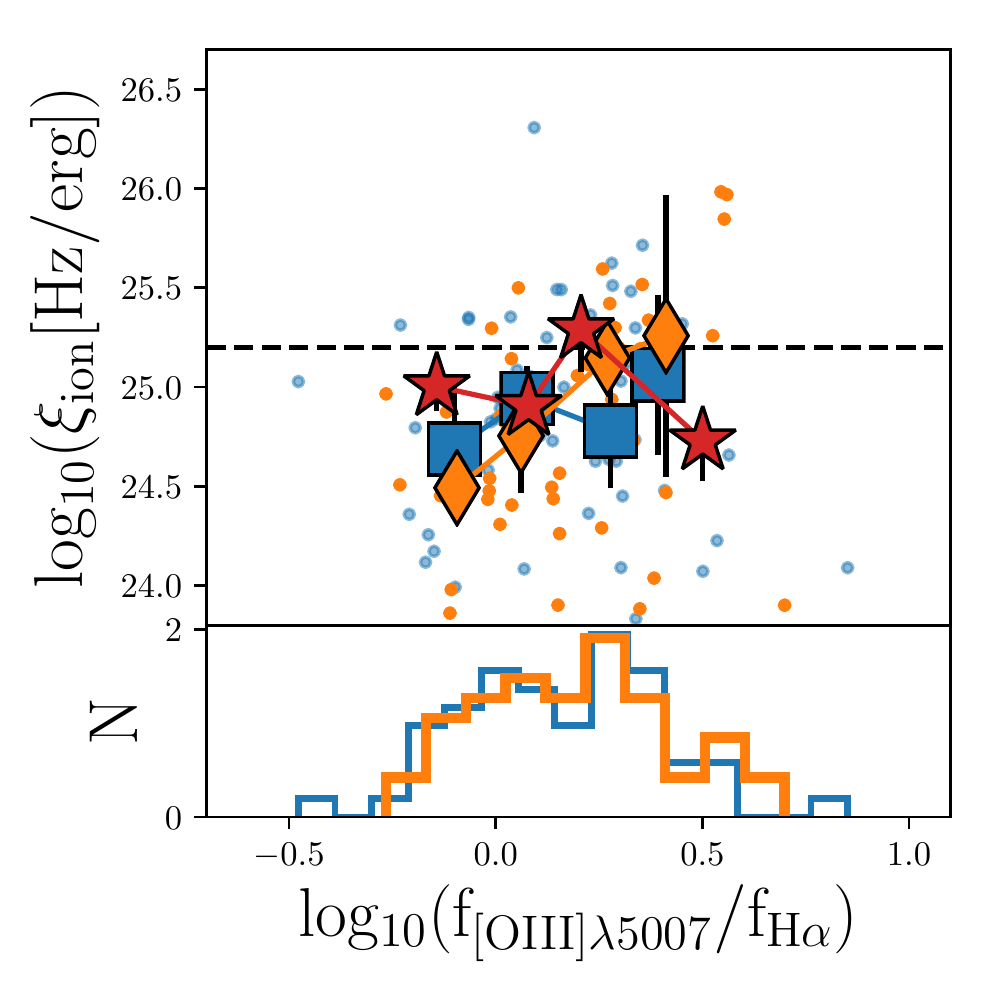}
\includegraphics[scale=0.6]{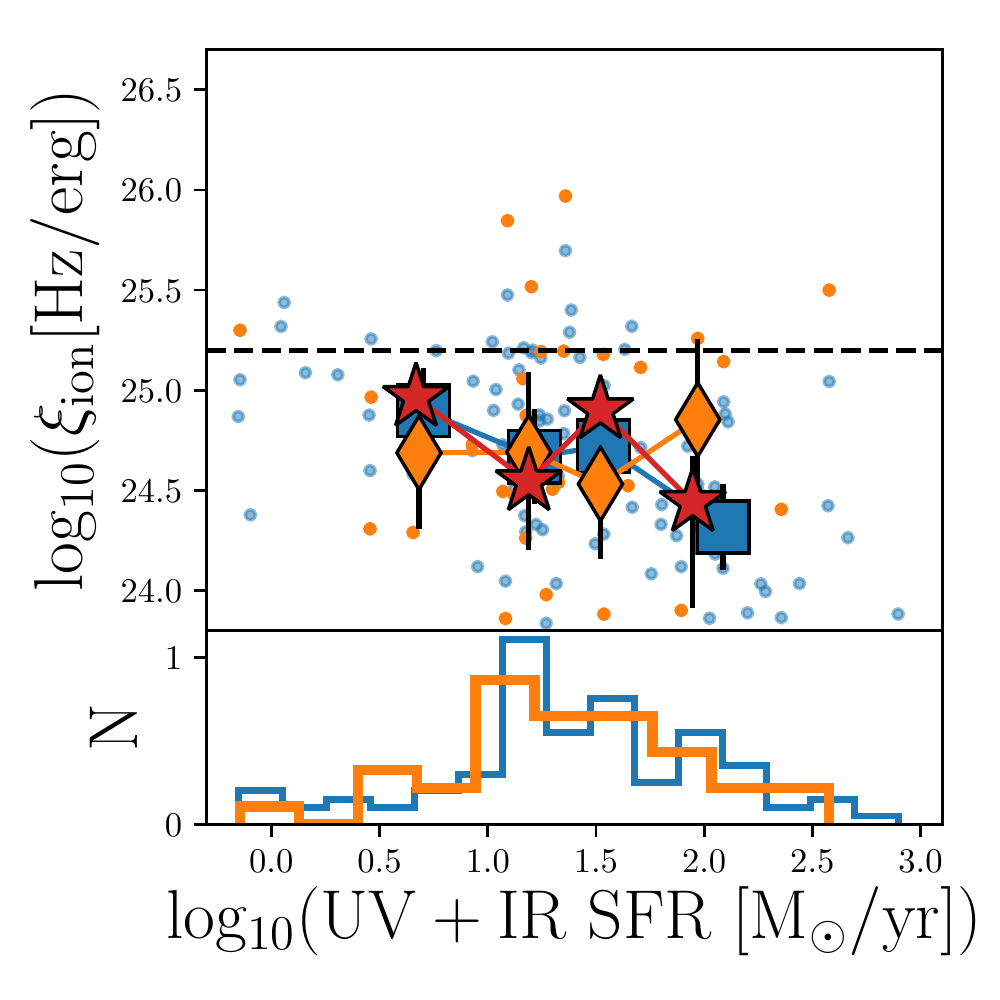}
\includegraphics[scale=0.6]{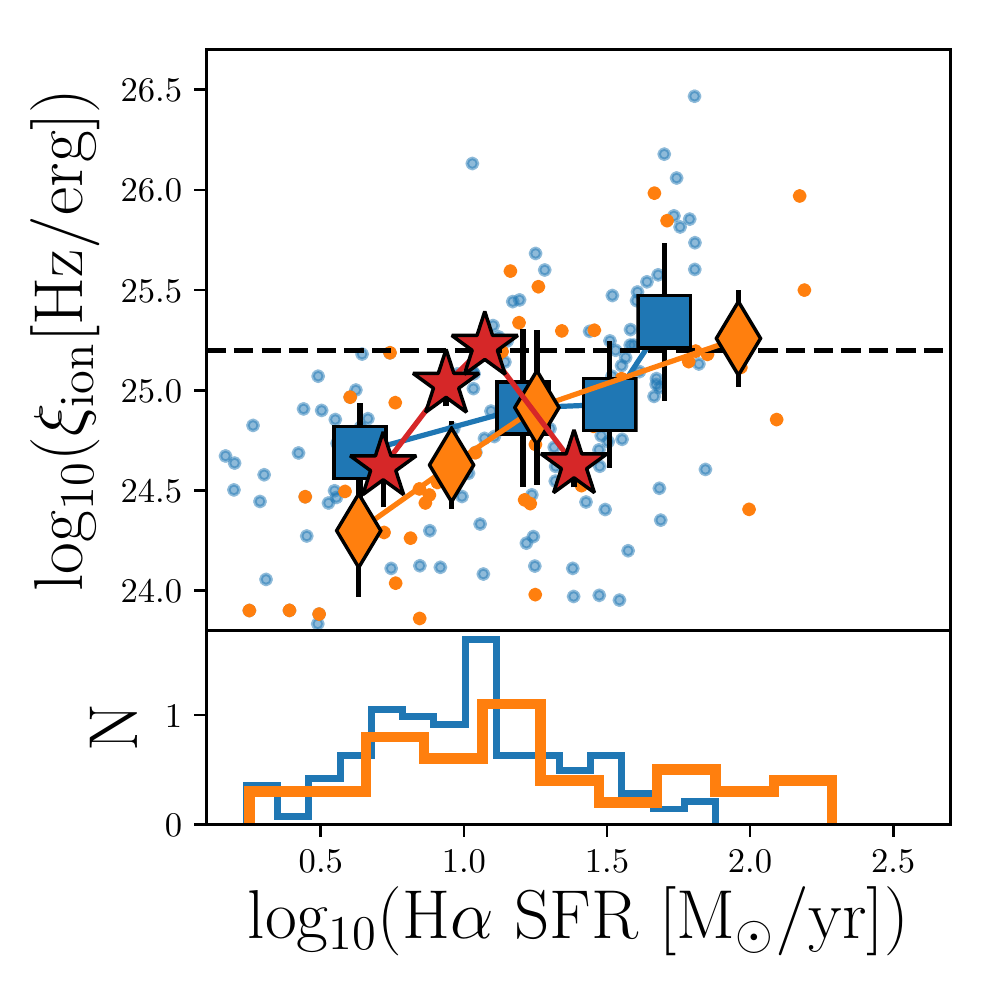}
\caption{The distribution of \xiion\ of our sample as a function of various galaxy properties: 
{\bf Top left:} UV continuum slope ($\beta$), {\bf top centre:} UV magnitude, {\bf top right:} stellar mass, 
{\bf bottom left:} \OIII/\Halpha\ ratio , {\bf bottom centre:} UV+IR SFR (only galaxies with at least one detection in photometric bands $>5\mu m$ in the observed frame are shown), and {\bf bottom right:} \Halpha\ SFR. 
We bin galaxies in each of the considered properties and illustrate the median value in each bin and the $1\sigma$ scatter parameterized by the median absolute deviation. 
We additionally stack set A galaxies with MOSFIRE $H$ band observations in four bins parametrized by the distribution in each of the galaxy properties and show the measurements computed for the stacks with associated errors computed using bootstrap resampling.  
A typical error bar for \xiion\ measurements of the individual galaxies is shown below the legend of the top left panel. 
The black horizontal dashed line shows \citet{Robertson2013} log$_{10}$(\xiion\ [Hz/erg])=$25.2$ value. 
The lower panels of the figures illustrate the distribution of the values in each of the parameters. 
\label{fig:xiion_vs_galaxy_params}}
\end{figure*}

In Figure \ref{fig:xiion_vs_galaxy_params} we show the distribution of \xiion\ as a function of various galaxy properties. Along with individual galaxies, we also bin in quartiles to show the median trend of \xiion\ with these galaxy properties. For each property, we stack galaxies with MOSFIRE H band observations in each of the quartiles to compute an average balmer decrement, which is used to correct for dust extinction of the nebular emission lines.

\paragraph{$\beta$} 
We observe a statistically significant moderate negative correlation between $\beta$ and \xiion\ (Spearman rank-order correlation coefficient \citep{Spearman1904} of $r_s, p_s=-0.6,1.3\times10^{-13}$ and $r_s, p_s=-0.5,2.4\times10^{-4}$ for set A and B, respectively).
Such a trend has also been observed at $z\sim2$ \citep[e.g.,][]{Shivaei2018} and at $z\sim4$ \citep[e.g.,][]{Bouwens2016c} but only for galaxies with $\beta<-2.0$, thus for galaxies with relatively low dust attenuation compared to our sample. 
The lowest $\beta$ observed for our sample is $\sim-2.3$. 
We rule out line flux detection levels as a cause because we expect galaxies with low $\beta$ to be highly star-forming dust free young systems. 

\paragraph{UV Magnitude} 
We compute the UV magnitude of our sample by integrating the FAST++ best-fit SED template in the rest-frame using a box car filter at $1500\pm175$ \AA. 
We do not find any evidence for any statistically significant correlations  ($r_s, p_s=-0.1,0.12$ and $r_s, p_s=0.03,0.83$ for Set A and B, respectively) between UV magnitude and \xiion. 
The stacked galaxies in UV magnitude bins also show a similar trend to the average trend of the individual galaxies. 
80\% of our set A galaxies lie at  UV magnitude $>-18.8$, thus we cannot constrain the evolution of \xiion\ with UV magnitude for galaxies with fainter UV magnitudes, which is expected to dominate UV luminosity function at $z>6$ \citep{Bouwens2015}. 
Whether \xiion\ show a UV magnitude dependence is still unclear, with some studies showing evidence for no correlation at $z\sim2-5$ \citep[e.g][]{Bouwens2016c,Shivaei2018,Lam2019a} and some showing evidence for a correlation \citet{Matthee2017}.

\paragraph{Stellar mass} 
Both Set A and B galaxies show evidence for a statistically significant moderate negative correlation of \xiion\ with stellar mass  ($r_s, p_s=-0.4,2.1\times10^{-6}$ and $r_s, p_s=-0.3,0.08$ for set A and B, respectively).
A similar distribution is also evident for our stacked sample based on balmer decrements computed using stellar mass stacks.
As shown by the stellar mass histograms, both set A and B galaxies show a similar distribution in stellar mass. 
We perform a two sample K-S test between the parent ZFOURGE sample used for target selection of ZFIRE and set A and B galaxies and find that we cannot rule out the null hypothesis that both set A and set B galaxies are sampled from the same parent population.  
Therefore, we rule out any selection effects based on stellar mass selection from ZFOURGE to play a role in driving these correlations. 

In terms of the average trends, the lowest mass galaxies show a slight enhancement in \xiion\ but the distribution flattens out at higher stellar masses and and shows a prominent decline at the highest stellar masses for both set A and B galaxies.
The ZFIRE \Halpha\ detected sample reaches an 80\% mass completeness at  $log(M_*/M_\odot)\sim9.3$ \citep{Nanayakkara2016} and set B also shows a similar completeness level. 
Therefore, we cannot make strong conclusions on the excess of \xiion\ in the lowest mass bin which has a median $log(M_*/M_\odot)\sim9.3$ \msol. 
Observations by \citet{Shivaei2018} and predictions by cosmological hydrodynamical simulations by \citet{Wilkins2016b} show a similar enhancement at lower stellar masses but is attributed to a secondary effect compared to stellar population properties.

\paragraph{ \OIII/\Halpha\ flux ratio} 
We select 58 galaxies with S/N$\geq3$ for the \OIII\ emission line from our set A sample. At $z\sim2$, \OIII\ and \Hbeta\ both fall in the MOSFIRE H band and all 49 galaxies in set B are also detected with \OIII. 
We compute the intrinsic \OIII\ line flux similar to \Halpha, namely using the \citet{Cardelli1989} dust law with the balmer decrement computed for the stacked galaxies in each of the quartiles. 
Set A galaxies show no evidence for a statistically significant correlation  ($r_s, p_s=0.06,0.6$), but set B galaxies do show a moderate positive correlation of \xiion\ with \OIII/\Halpha\ ratio  ($r_s, p_s=0.4,2\times10^{-3}$). 
The stacked galaxies also show a flat distribution, however, the lowest \OIII/\Halpha\ bin shows a decline in \xiion. 
 
\Halpha\ luminosity traces the young ionizing stars in a galaxy while the conversion factor between the number of high-energy photons and the \OIII\ luminosity is strongly dependent on the metallicity and the ionizing parameter \citep[e.g.,][]{Kewley2013a}. 
The observed correlation for set B suggests that galaxies with higher ionization parameters tend to have higher \xiion\ similar to observations by \citet{Shivaei2018}, however, we cannot make strong conclusions due to the absence of a correlation for set A galaxies.

\paragraph{SFR}
Both, \Halpha\ emission line luminosity and UV luminosity are direct traces of star-formation with different age dependencies \citep[e.g.,][]{Haydon2018}, thus, we expect \xiion\ to show some correlation with the SFR. 
We select 98 galaxies from set A with detections in at least one of the \emph{Spitzer}/MIPS or \emph{Herschel}/PACS bands and use ZFOURGE UV+IR SFRs \citep{Tomczak2016} to investigate dependencies of \xiion\ with SFR. 
All set B galaxies satisfy the above criteria. 
Our set A sample shows a moderate statistically significant negative correlation of \xiion\ with UV+IR SFR ($r_s, p_s=-0.5,9.1\times10^{-7}$), while set B does not show any evidence for a statistically significant correlation ($r_s, p_s=0.08,0.65$). 

In terms of \Halpha\ SFRs, both set A and B galaxies show a statistically significant moderate positive correlation with \xiion\ ($r_s, p_s=0.5,6.9\times10^{-10}$ and $r_s, p_s=0.6,6.3\times10^{-7}$ for set A and B, respectively). 
The stacked galaxies show a similar positive trend, however, the highest SFR bin shows a decline in SFR, which is primarily driven by $\sim0.2$ dex increase in UV luminosity in the highest SFR bin. 
Given both set A and B galaxies show similar distributions for UV+IR and \Halpha\ SFRs, it is unlikely that a bias in SFR would drive the different observed correlations between UV+IR and \Halpha\ SFRs.

Our analysis of \xiion\ with with various galaxy observables/properties demonstrate that our sample does not show any strong trends with variables that are commonly used for selection (e.g., stellar mass, $\mathrm{M_{UV}}$).
The observed trends of \xiion\ are moderate at most and also agrees well with other studies where available. 
We conclude that both set A and B samples are relatively unbiased samples of star-forming galaxies at $z\sim2$.


\subsection{Combining \xiion\ with \Halpha\ EW and optical colors}
\label{sec:xi_ion_and_Halpha_EW}

\xiion, \Halpha\ EW, and rest-frame optical colors are diagnostics of specific SFR (sSFR) sensitive to different stellar masses. 
As shown by Equation \ref{eq:N_H},  $N(H)\propto$ \Halpha\ and therefore is sensitive to young O type stars with masses $\gtrsim20$ \msol. 
The continuum at \Halpha\ of a star-forming galaxy is dominated by red giant stars with masses $\sim0.7-3$ \msol, while UV luminosity trace O and B type stars with masses $\gtrsim3$\msol. 
Therefore, \Halpha\ EW traces the ratio of the short lived massive O stars to the older red giants stars, while \xiion\ traces the ratio of massive O type stars to less massive O and B type stars and are sensitive to the mass distribution in a stellar population at different parts of the IMF.

Given that the mass of the stars determine its main-sequence life-time, \Halpha\ EW and \xiion\ are both also sensitive to the age from the most recent star-burst in a stellar population.  
The \boxfil\ color \citep[box car filters at $3400\pm150$ \AA\ and $5500\pm150$ \AA\ chosen to avoid spectral regions with strong emission lines; see Appendix B in][for further details]{Nanayakkara2017} is sensitive to the ratio of bluer stars to redder stars. 
Thus \boxfil\ color, \Halpha\ EW, and \xiion\ are also sensitive to the SFH/age of a stellar population.

In this section we combine the analysis of \Halpha\ EW and \boxfil\ color with \xiion\ to investigate whether we can make stronger constraints on the nature of the stellar populations.

The \Halpha\ EW and rest-frame optical colors of star-forming galaxies have been studied in detail as a tracer of high mass stellar IMF and of stellar rotation and binaries in stellar populations \citep[e.g.,][]{Kennicutt1983,Hoversten2008,Gunawardhana2011,Nanayakkara2017}. 
From our full sample, we remove galaxies with multiple objects within the MOSFIRE slits or galaxies that had bright sources close to the slit edges and select 77 (out of which 31 are in Set B) galaxies to compute the \Halpha\ EW using ZFOURGE $Ks$ band photometry. 
We remove the \Halpha\ flux contribution from the photometric flux and compute a continuum flux assuming that other emission lines within $Ks$ band to have a negligible contribution to the total photometry.  
We then approximate \Halpha\ EW as the fraction between \Halpha\ line flux and the continuum estimated from the photometry. 
In order to investigate any systematic offsets in computing \Halpha\ EWs using ZFOURGE photometry, we select a subsample of 38 galaxies with confident $K$ band continuum detections in MOSFIRE spectra \citep{Nanayakkara2017} and compare the difference in \Halpha\ EW. 
We find a good agreement between \Halpha\ EWs computed using spectroscopically to photometrically determined continuum levels with a median $\mathrm{\Delta\log_{10}(EW)=-0.02\pm0.11\AA}$. 
Additionally, we note that all ZFIRE spectra are corrected for slit loss using broadband photometry from $HST\ F160W$  and FourStar $Ks$ band fluxes \citep{Nanayakkara2016}.


\subsubsection{Simple parametric SFHs using BPASS stellar population models}
\label{sec:parametric_sfhs}

\begin{figure*}
\includegraphics[trim = 8 0 5 5, clip,  scale=0.625]{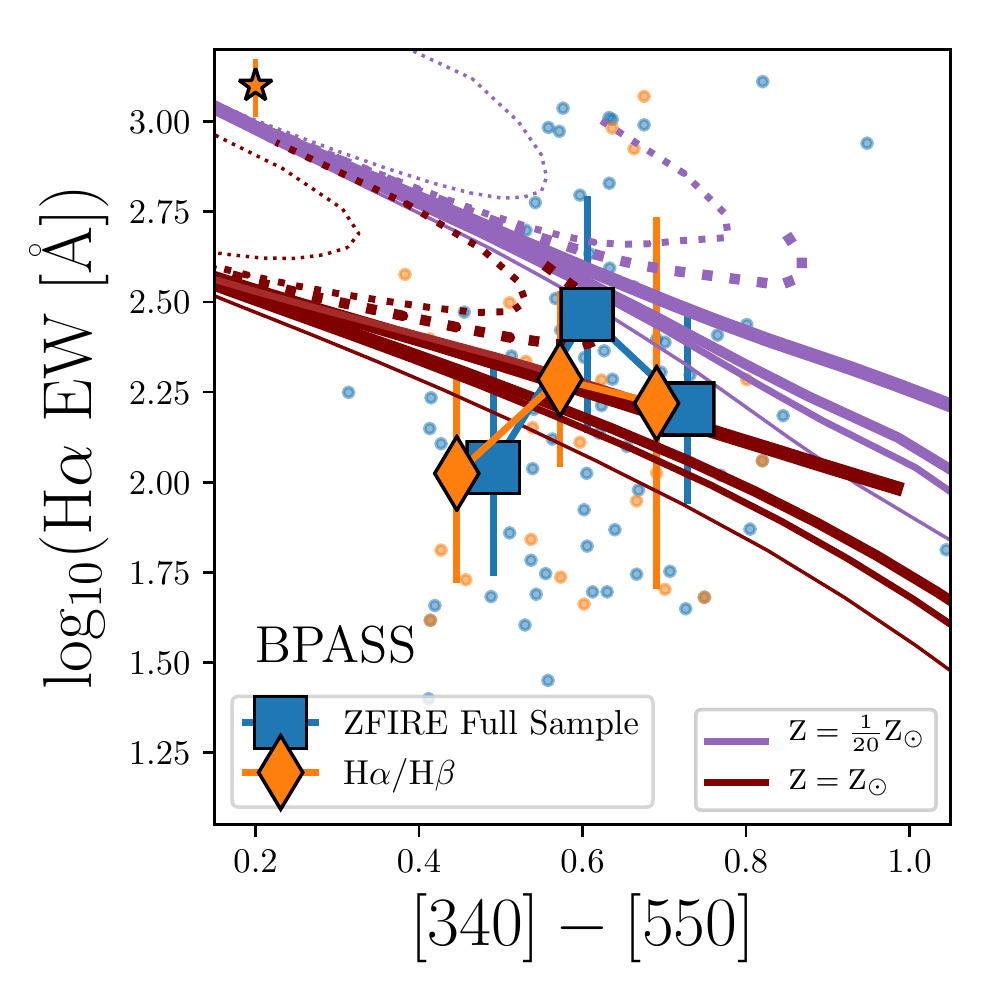}
\includegraphics[trim = 10 0 5 5, clip, scale=0.625]{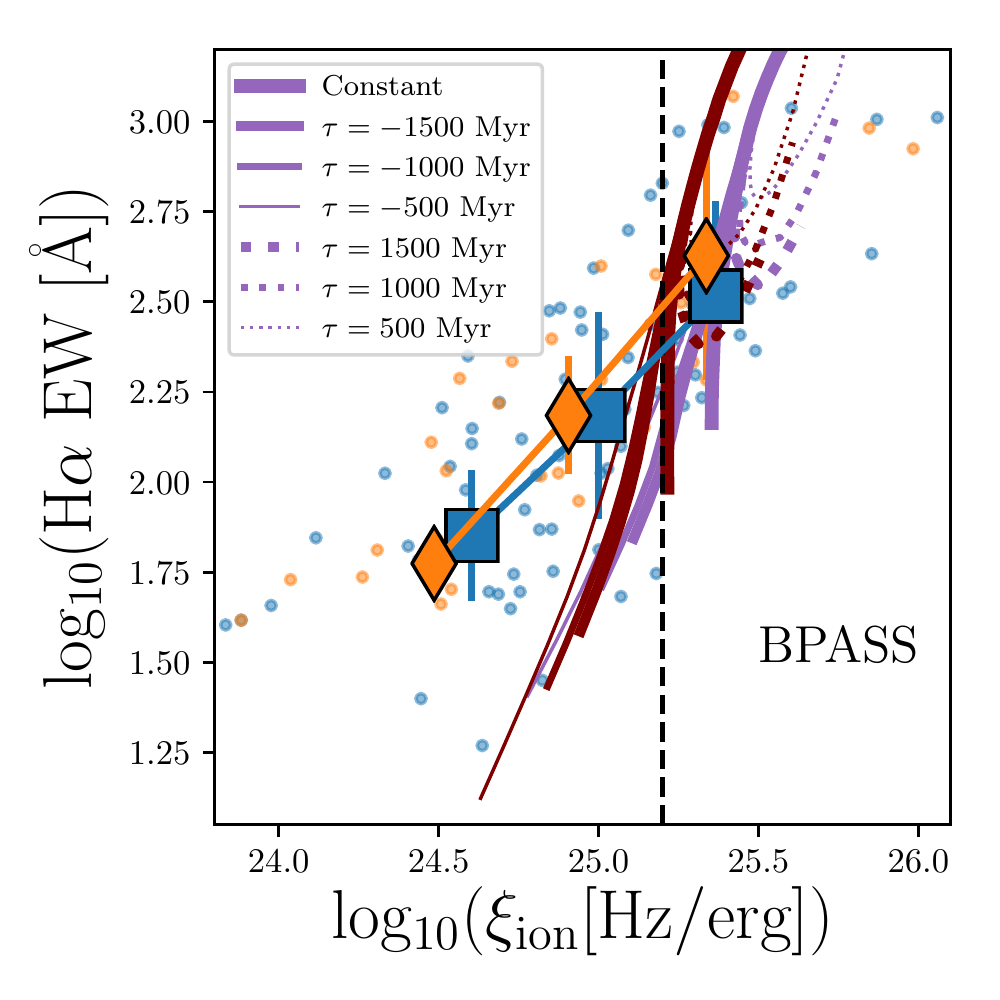}
\includegraphics[trim = 10 0 5 5, clip, scale=0.625]{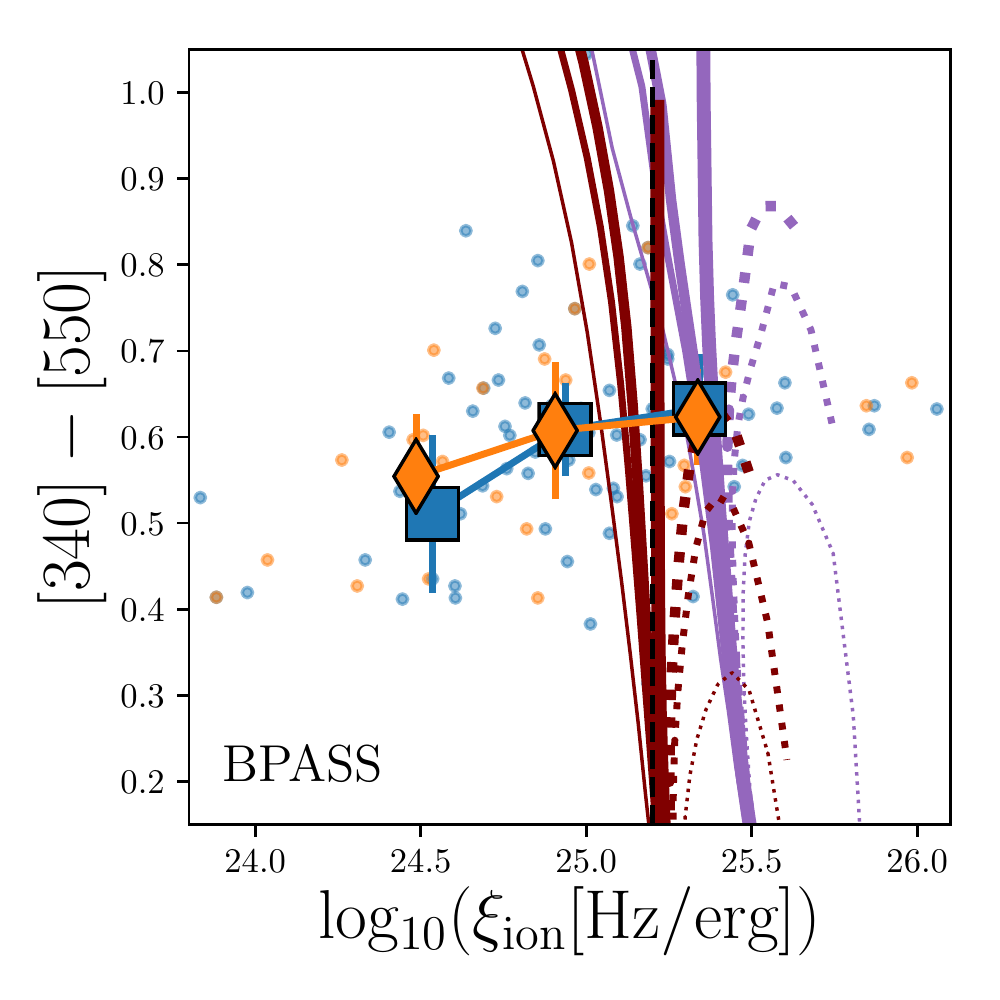}
\caption{ {\bf Left:} \Halpha\ EW as a function of dust corrected \boxfil\ color, 
{\bf center:} \Halpha\ EW, and {\bf right:} dust corrected rest-frame optical color (\boxfil) as a function of \xiion. 
Galaxies are binned in equal number bins in the x axis with the scatter parameterized by the median absolute deviation. 
We overlay stellar population models from BPASSv2.2.1 for constant SFH models (1\msol/yr) and exponentially increasing and decreasing SFHs with $\tau=1500, 1000, 500$ Myr for \zsol\ and 1/20 \zsol\ metallicities. All models are computed for a \citet{Salpeter1955} like $\Gamma=-1.35$ slope IMF for stellar masses in the range of $0.5-300$\msol\ and a $\Gamma=-0.3$ slope for masses in the range of $0.1-0.5$\msol. All models terminate at $t\sim3100$ Myr which is the age of the Universe at $z\sim2$. 
The largest error bar for a single galaxy in the \Hbeta\ detected sample is shown by the top left star in the top left panel.  
The observed distribution of our galaxies in \Halpha\ EW and rest-frame optical color space is well reproduced by the BPASS models by varying the SFH and the stellar metallicity, however, the predicted \xiion\ values are consistently too high for the observed \Halpha\ EW and rest-frame optical colors 
\label{fig:ew_color_xiion}}
\end{figure*}


We first use BPASSv2.2.1 binary stellar population models with simple parametric SFHs to compare with the distribution of our observed sample in \xiion, \Halpha\ EW, and \boxfil\ color space. 
In Figure \ref{fig:ew_color_xiion} we show the \Halpha\ EW vs \boxfil\ color of our sample and expectations from BPASS stellar population models with \zsol\ and 1/20th \zsol\ metallicities. 
The models are computed for constant and exponentially increasing and decreasing SFHs with a \citet{Salpeter1955} like IMF. 
Galaxies with high \Halpha\ EW for a given \boxfil\ color prefers lower metallicity tracks compared to galaxies with low \Halpha\ EWs. The average distribution of galaxies in \Halpha\ EW and \boxfil\ color can be explained by the BPASS models. 
However, we note that there is a fraction of galaxies with lower \Halpha\ EWs and/or bluer optical colors than what is expected from the BPASS models. Including effects of random star-bursts over smooth SFHs in stellar population models could explain this subset of galaxies \citep{Nanayakkara2017} and we discuss this further in Section \ref{sec:star_bursts}.

In Figure \ref{fig:ew_color_xiion} we also show the distribution of \Halpha\ EW and \boxfil\ color as a function of \xiion. 
In terms of \Halpha\ EW there is a statistically significant observed trend, where galaxies with higher \Halpha\ EWs show higher \xiion\ values. 
This trend is expected because both axes trace the number of hydrogen ionizing photons in the nominator and therefore are correlated with each other. 
In order to verify that low S/N in the \Halpha\ measurement does not lead to the observed correlation, we compute the Spearman's rank correlation coefficient for galaxies with \Halpha\ S/N$>10$ and find that the statistically significant trend of \Halpha\ EW with \xiion\ still holds.

In Figure \ref{fig:ew_color_xiion} center and right panels we also show the same BPASS models that well described the observed distribution of \Halpha\ EW and \boxfil\ color of our sample. 
The observed galaxies on average show higher \Halpha\ EWs for a given \xiion, specially for galaxies with $\log_{10}$(\xiion [Hz/erg]) $<25.0$. 
The models diverge from the data at \Halpha\ EW $\lesssim2.25$\AA\ in \Halpha\ EW vs \xiion\ space, while in \Halpha\ EW and \boxfil\ color space models only diverge from the data at \Halpha\ EW $\lesssim2.0$\AA.

\xiion\ vs \boxfil\ color  (Figure \ref{fig:ew_color_xiion}) shows evidence for a statistically significant moderate trend, where galaxies with higher \xiion\ show slightly redder optical colors. 
This suggest that in light weighted terms optical colors of the high \xiion\ sample may be dominated by the older stellar populations.  
Therefore, if galaxies with high \xiion\ does harbor star-bursts, the relative strength of the star-burst compared to the past SFH should be low. 
Within the BPASS parametric SFHs explored in Figure \ref{fig:ew_color_xiion}, our observed galaxies with $\log_{10}$(\xiion [Hz/erg]) $<25.0$ are bluer compared to the BPASS models.

BPASS model tracks show a strong dependence on Z in \Halpha\ EW vs \boxfil\ color space. At all times, the low-Z models has higher \Halpha\ EWs compared to the \zsol\ models, however, low-Z models evolve fast in the \boxfil\ colors to be redder. 
The observed distribution of our galaxies are in general well explained by BPASS models the \Halpha\ EW vs \boxfil\ space by simply varying the Z and the exponential decay time scale of the SFH.

In \Halpha\ EW vs \xiion\ and \boxfil\ vs \xiion\ space, BPASS models do not well represent the observed data. 
The drop in \Halpha\ EW for a given \xiion\ was too high in the BPASS models in order to match with the observed data. Additionally, model galaxies were too red at $log_{10}(\xi_{ion}[Hz/erg])\lesssim25.0$. 
Since the observed distribution in \Halpha\ EW vs \boxfil\ space is matched well by the BPASS models, it seems likely that the balance between the production rate of hydrogen ionizing photons and the UV luminosity drives the discrepancy between the models and data. 
If BPASS models with parametric SFHs are to match with the observed distribution of galaxies, UV luminosity at fixed \Halpha\ flux should decrease, thereby increasing the \xiion. 

Can the discrepancy between the BPASS models and our observations be resolved by introducing more complex SFHs with star-bursts?  
Given that our sample traces the typical star-formation stellar-mass relation for $z\sim2$ star-forming galaxies \citep{Nanayakkara2016,Nanayakkara2017} of \citet{Tomczak2016}, we expect simple parametric SFHs on average to be an accurate description of the SFHs of our systems. 
However, at $z\sim2$ galaxies are at the peak of the star-formation rate density and therefore it is likely that a fraction of our galaxies may in fact be in a star-burst phase. 
Next, we investigate the behavior of star-bursts in \xiion, \Halpha\ EW, and \boxfil\ color space.


\subsubsection{Star bursts using Starburst99 stellar population models}
\label{sec:star_bursts}

\begin{figure*}
\includegraphics[trim = 0 0 0 0, clip, scale=0.5]{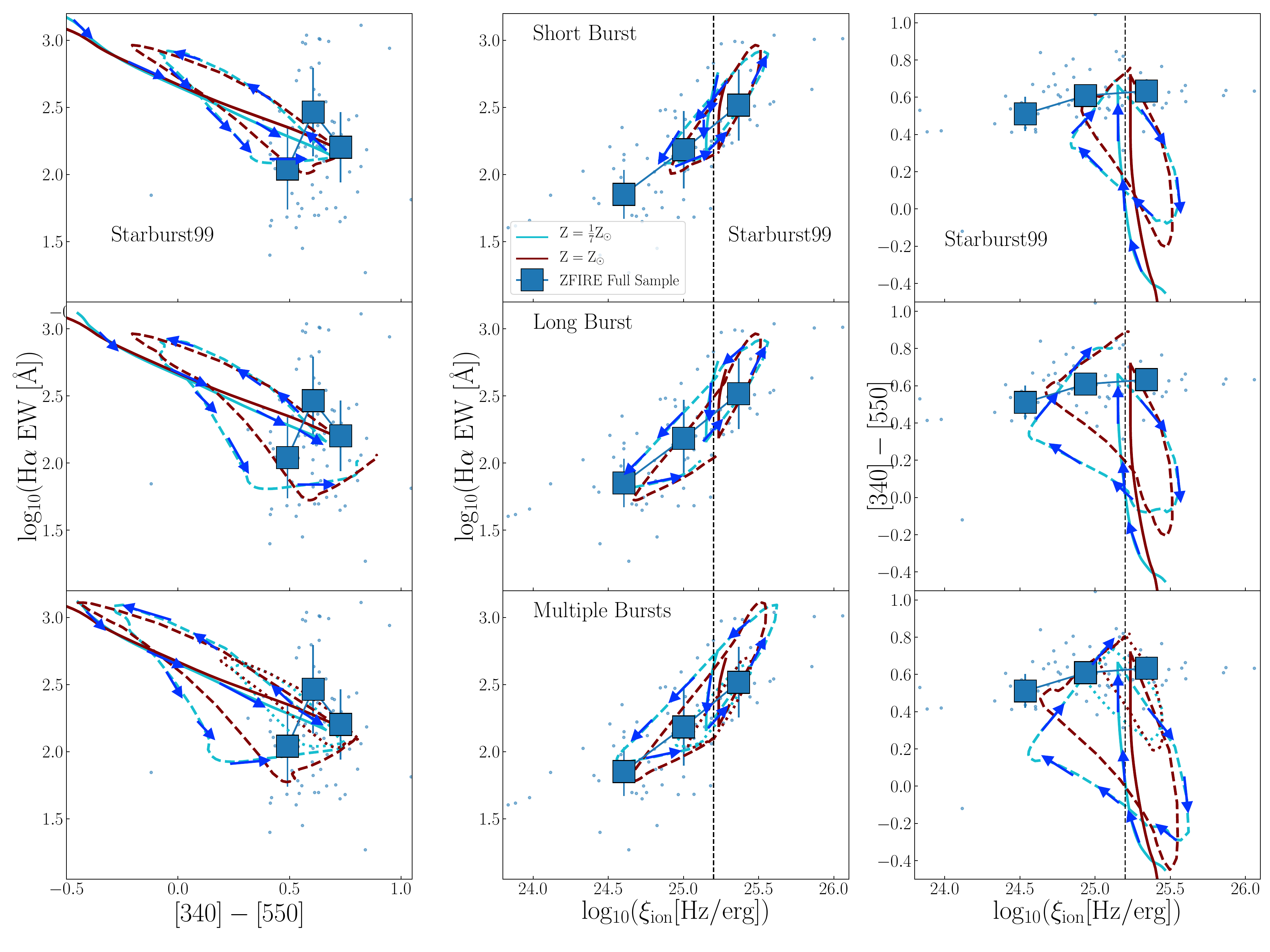}
\caption{ The behavior of starbursts in {\bf left:}  \Halpha\ EW vs \xiion, {\bf center:} \boxfil\ color vs \xiion, and {\bf right:} \Halpha\ EW vs \boxfil\ color space using the Starburst99 stellar evolution code with Geneva stellar tracks that account for effects of stellar rotation \citep{Leitherer2014}. 
From top to bottom the panels show; {\bf top:} a single star burst of $\times10$ the current SFR at $t=2500$Myr for $\Delta t=10$Myr, {\bf center:} a single star burst of $\times10$ the current SFR at $t=2500$Myr for $\Delta t=100$Myr, and {\bf bottom:} two star bursts of $\times20$ and  $\times5$ the current SFR at $t=2500$Myr and $t=2800$Myr for durations of $\Delta t=20$Myr and $\Delta t=10$, respectively. 
Model tracks up to the point of the star-burst is shown by the solid lines and after the onset of the star-burst the tracks are shown by the dashed lines.  If there is a secondary burst, the tracks after the onset of the secondary burst are shown by dotted lines. Arrows show how the model tracks evolve with time and is only shown for the 1/7 \zsol\ tracks.  
All models are normalized to the total mass formed by a 1\msol/yr constant SFH model at $z\sim2$. 
Only set A galaxies are shown in the figure to improve the clarity. 
The observed distribution of galaxies in this space can be explained well if most galaxies are in a longer duration star-burst/post star-burst phase. 
\label{fig:ew_color_xiion_models}}
\end{figure*}


We use Starburst99 models with Geneva stellar tracks that incorporate effects of stellar rotation in stellar evolution \citep{Leitherer2014} to analyze the effect of star bursts. We switch from BPASS to Starburst99 models for this analysis since we are able to perform finer time sampling at 1 Myr intervals in Starburst99, which is crucial to finely track the effect of star-bursts.

\begin{deluxetable}{lccc}
\tabletypesize{\footnotesize}
\tablecaption{ Starburst99 star-burst parameters used in the analysis
\label{tab:star-burst_properties}}  
\tablecolumns{5}
\tablewidth{0pt} 
\startdata
\hline \hline \\ 
Name & Burst Time\tablenotemark{a} 	    & Burst Strength & Burst Length  \\
	 & 			  (Myr) &  				 & 			(Myr)  \\

\hline \\ 
Short  		& 2500	& 	$\times10$		& 	10 	 \\
Long  		& 2500	& 	$\times10$		&  100	 \\
Multiple  	& 2500	 	& $\times20$	&  	20	 \\
		  	& 2800	 	& $\times5$		&  	10	 \\
\hline \\ 
Simulations\tablenotemark{b} & 	$$200-3000$$		& 	$\times5-100$			& $10-100$		\\
\tablenotetext{a}{Defined from the onset of star-formation.}
\tablenotetext{b}{$2-10$ bursts are chosen randomly within these parameters to construct the SFH.}
\end{deluxetable}

In Figure \ref{fig:ew_color_xiion_models} we investigate three different burst scenarios with varying burst strengths and burst lengths overlaid on constant SFH models. 
We tune the burst strengths and lengths to produce SFHs that cover the observed \xiion, \Halpha\ EW, and \boxfil\ color space and our burst properties are in agreement with FIRE simulation predictions of star-bursts \citep{Sparre2017}.  
A summary of these burst properties is provided in Table \ref{tab:star-burst_properties}.

Short star-bursts in the post star-burst phase are able to maintain the observed high \Halpha\ EW of the galaxies while maintaining a $\log_{10}$(\xiion [Hz/erg]) $\gtrsim 24.8$. 
However, such bursts fail to reproduce the observed redder colors of the galaxies. 
Long lived bursts produce post star-burst tracks that could explain a majority of galaxies with low \xiion\ that have relatively low \Halpha\ EWs. 
Once multiple bursts are invoked in the last $\sim600$ Myr of the SFH of the galaxies, galaxies in the post star-burst phase show a similar behavior to the individual burst case. 
By invoking star-bursts with varying strengths and lengths the observed distribution at $z\sim2$ could be reproduced.

\begin{figure*}
\includegraphics[trim = 10 0 0 0, clip, scale=0.6]{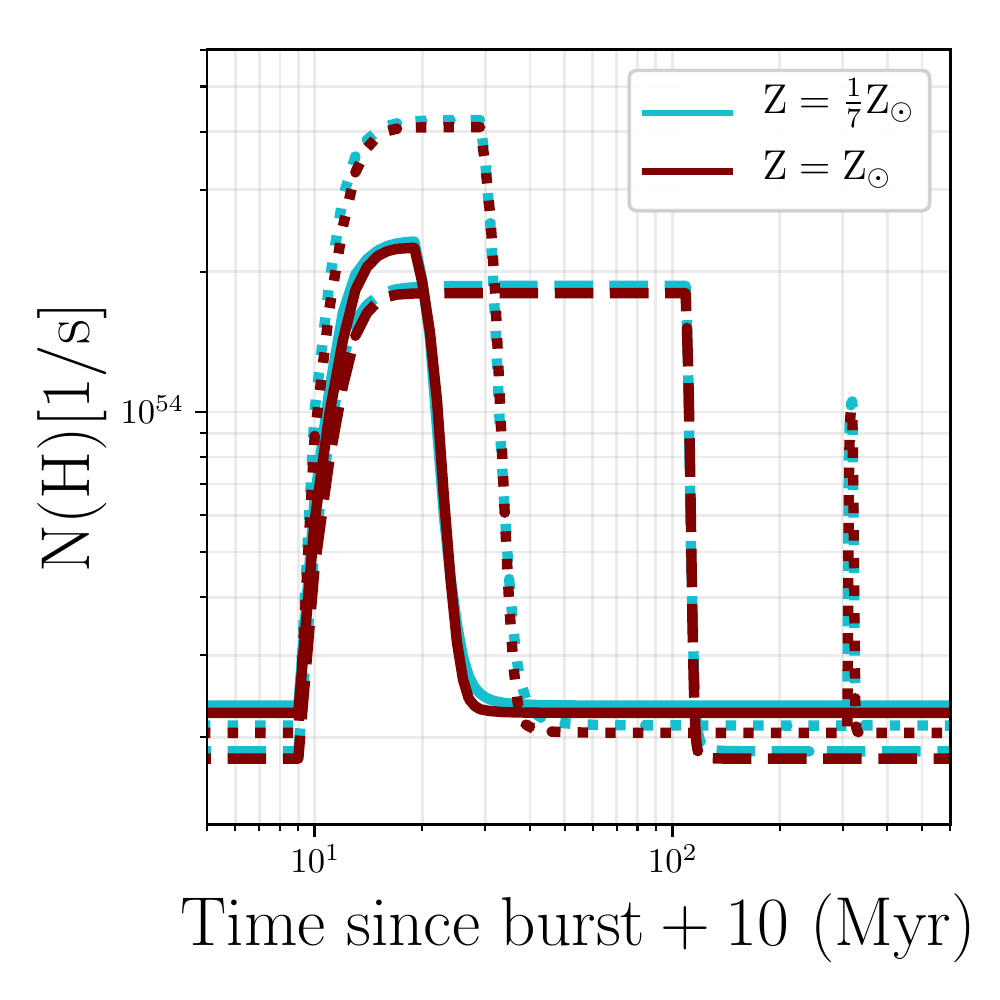}
\includegraphics[trim = 10 0 0 0, clip, scale=0.6]{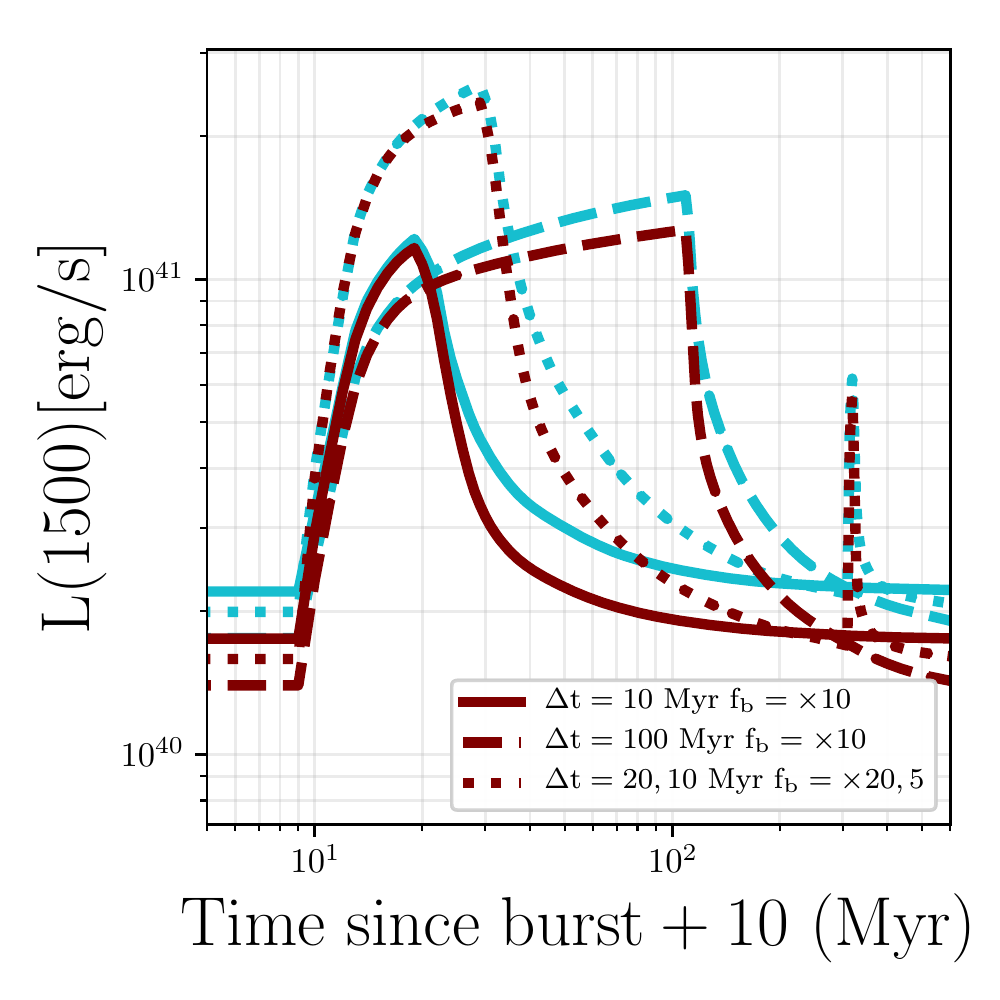}
\includegraphics[trim = 10 0 0 0, clip, scale=0.6]{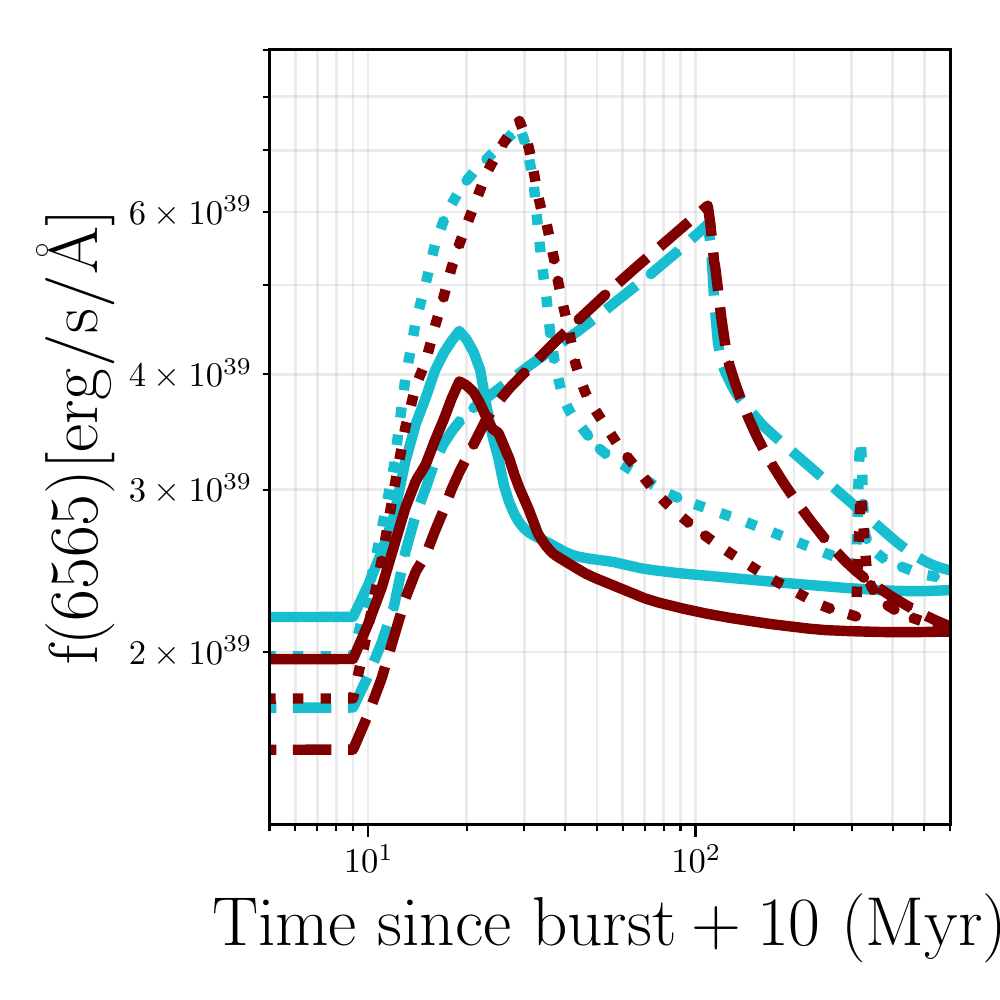}
\caption{ The time evolution of {\bf left:} the hydrogen ionizing photon production rate, {\bf center:} the UV luminosity at 1500\AA, {\bf right:} the continuum flux at 6565\AA\ of Starburst99 models computed using Geneva stellar tracks that include effects of stellar rotation.  
Models are shown for three different burst scenarios with different burst strengths and lengths identical to Figure \ref{fig:ew_color_xiion_models}. 
In order to improve the clarity in time evolution, models are shown from $t-10$ Myr from the burst.  
Introducing bursts have a strong influence on all three variables, however, effects are spread over different time-scales.     
\label{fig:bursts_time_evolution_1}}
\end{figure*}

\begin{figure*}
\includegraphics[trim = 10 0 5 5, clip, scale=0.625]{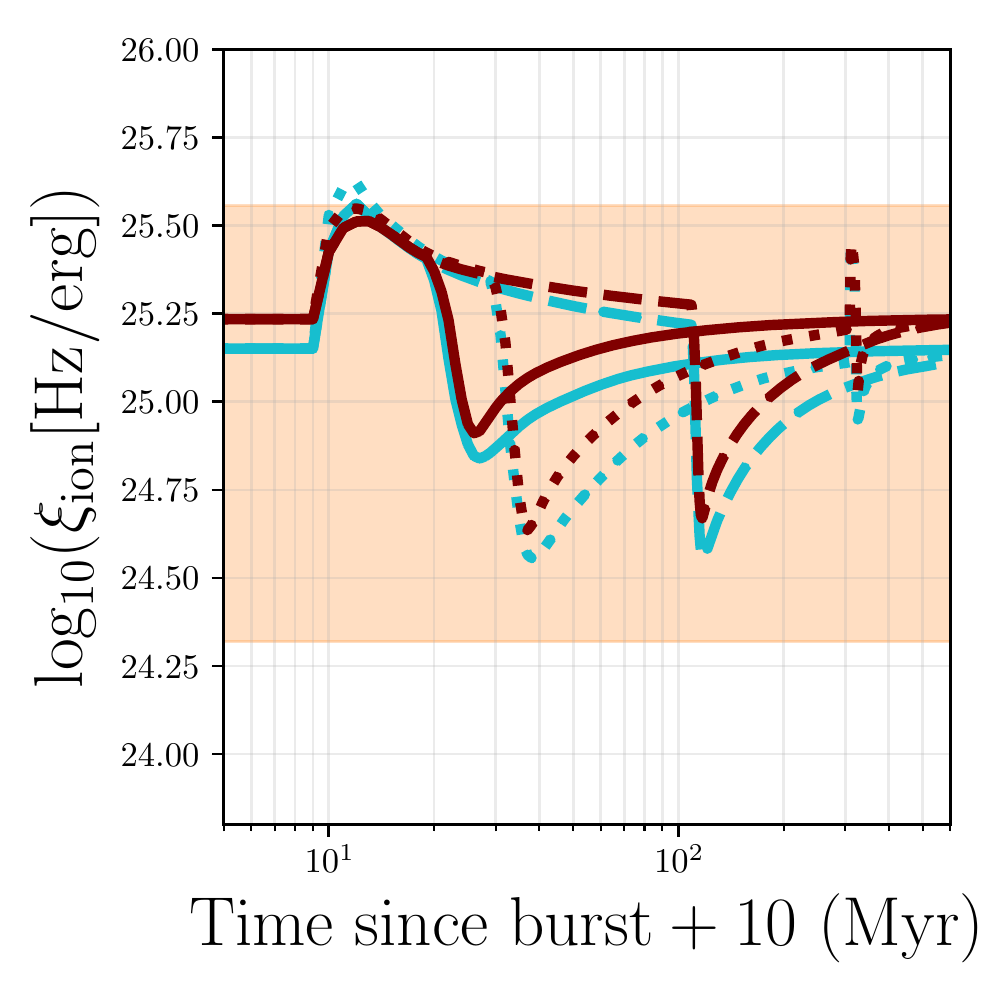}
\includegraphics[trim = 10 0 5 5, clip, scale=0.625]{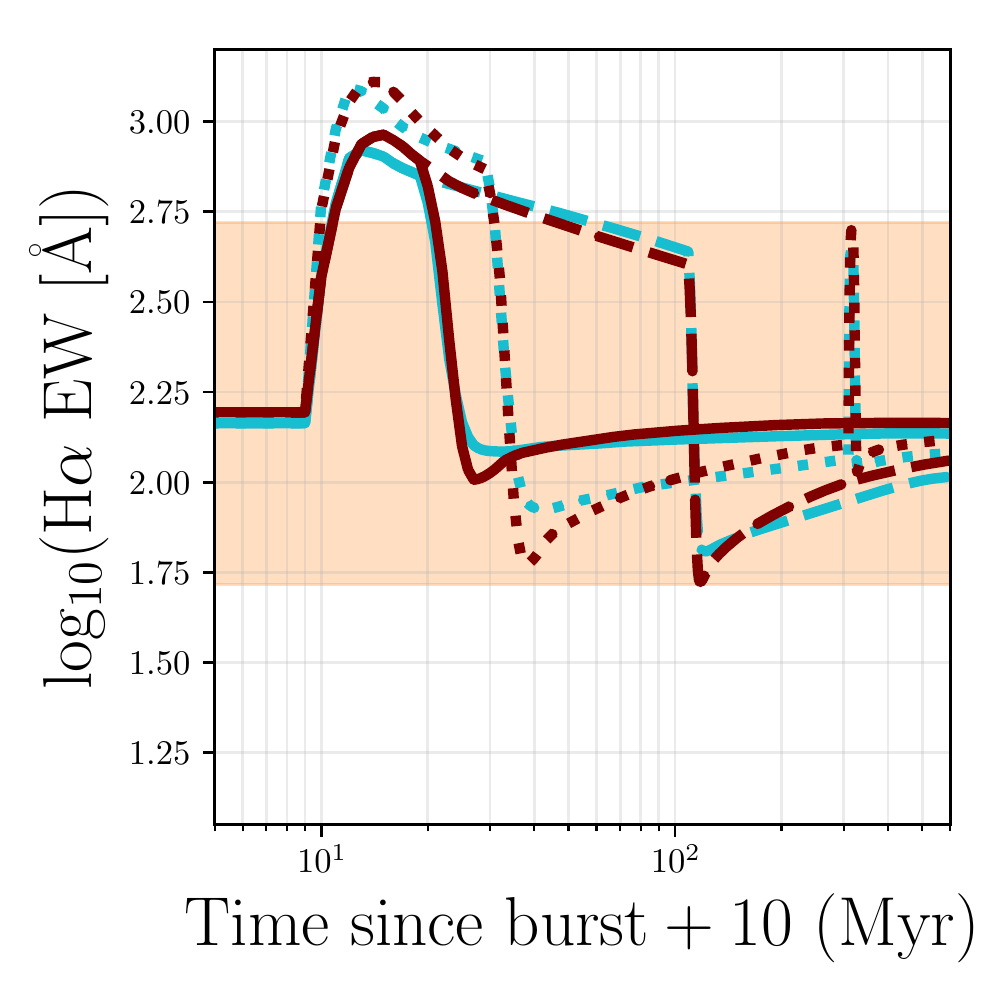}
\includegraphics[trim = 5 0  5 5, clip, scale=0.625]{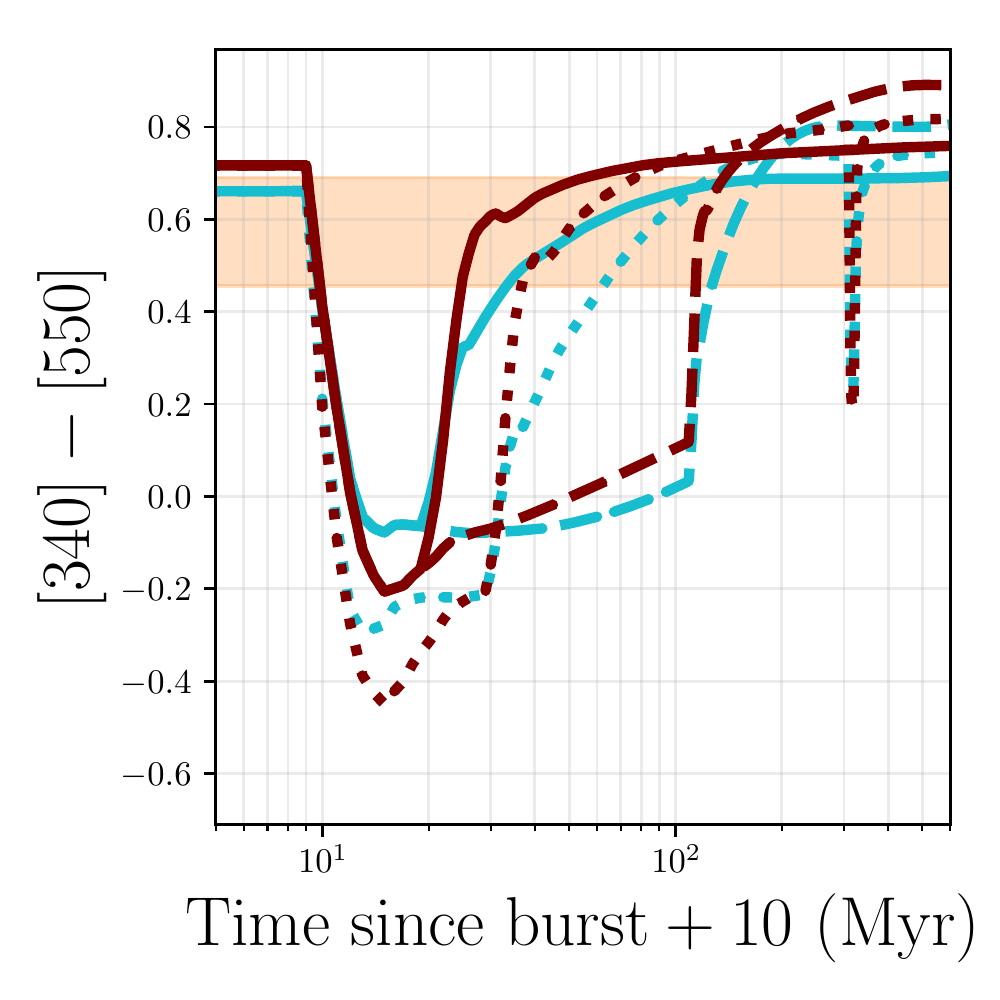}
\caption{ The time evolution of  {\bf left:} \xiion, {\bf center:} \Halpha\ EW, {\bf right:} \boxfil\ colors of Starburst99 models shown by Figure \ref{fig:bursts_time_evolution_1}. 
For each variable, the observed median$\pm 1\sigma$ distribution of the $z\sim2$ set A sample is shaded in yellow. 
The observed distribution of the galaxies can be reproduced by star-bursts.   
\label{fig:bursts_time_evolution_2}}
\end{figure*}

In Figures  \ref{fig:bursts_time_evolution_1} and \ref{fig:bursts_time_evolution_2}, we further investigate the effect of star-bursts and the relative time-scales on which our observed parameters change.  
At the onset of the star-burst, driven by the increase in the H ionizing photon production rate, the \Halpha\ EW and \xiion\ increase rapidly to their maximum values within $\sim3$Myr. 
UV luminosity takes $\sim10$ Myr to stabilize during the star-burst, thus, once a maximum \xiion\ is achieved at $\sim5$ Myr, \xiion\ starts to drop gradually due to the increase in the UV luminosity. 
In the post star-burst phase, the drop in the H ionizing photon rate occurs very rapidly within the typical life-time of massive O type stars of $\sim10$ Myr. 
Less massive O and B type stars that contribute to the UV luminosity are longer lived in the main sequence, thus UV luminosity only reach pre-burst levels $\sim100$ Myr after the star-burst. 
Therefore, after the end of the star-burst \xiion\ drops to a minimum and gradually rises up to the pre-burst levels driven by the reduction in UV luminosity.

The continuum at 6565\AA\ is dominated by red giant stars and therefore from the onset of the star-burst the continuum flux gradually increase to a maximum until the end of the star-burst. In the post-star burst phase, the continuum drops gradually and only reaches pre-burst continuum levels few $100$ Myr after the end of the star-burst. 
Thus, \Halpha\ EW show a similar time evolution to \xiion, however, in the post star-burst phase  \Halpha\ EW takes a longer time to reach pre-burst levels.

The \boxfil\ color of the galaxies is also very sensitive to star-bursts. 
Due to the massive blue stars formed by the star-burst, galaxies show an almost instantaneous shift to blue colors at the onset of the star-burst. 
Driven by the increase in post main-sequence redder stars, the \boxfil\ color gradually declines during the star-burst until the end of the star-burst. Galaxies turn redder within a very short time-scale in the post star-burst phase, where stronger/longer lived bursts show redder colors in the post star-burst phase compared to weaker/shorter lived star-bursts.

Within the context of Starburst99 models, we find metallicity to only have a weak influence on \xiion, \Halpha\ EW, and \boxfil\ colors. 
The hydrogen ionizing photon production rate of the Geneva rotational models only increase by $\times \sim1.04$ between \zsol\ to 1/7th \zsol\ models. 
The strongest influence of metallicity is on UV luminosity, where lower metallicity stars show $\sim25\%$ higher UV luminosity compared to higher metallicity stars. This is possibly driven by stellar rotation, where higher metallicity stars loose angular momentum faster due to their optically thick winds.

Starburst99 models with star-bursts could reproduce our observed distribution of $z\sim2$ galaxies in \xiion, \Halpha\ EW, and \boxfil\ space. 
However, in order to satisfy the observables, a majority of our galaxies should lie in a post star-burst phase and the time-window on which the models populate the observed space is short compared to the total age of the Universe at $z\sim2$. 
We generate 1000 Starburst99 model galaxies with a constant SFH and overlay multiple bursts with randomly selected strengths and lengths at random times in its SFH and perform 10,000 bootstrap samples from the model grid between $1500-3100$ Myr time window. 
A summary of the burst properties is also presented in Table \ref{tab:star-burst_properties}.

\begin{figure*}
\includegraphics[trim = 0 0 0 0, clip, scale=0.6]{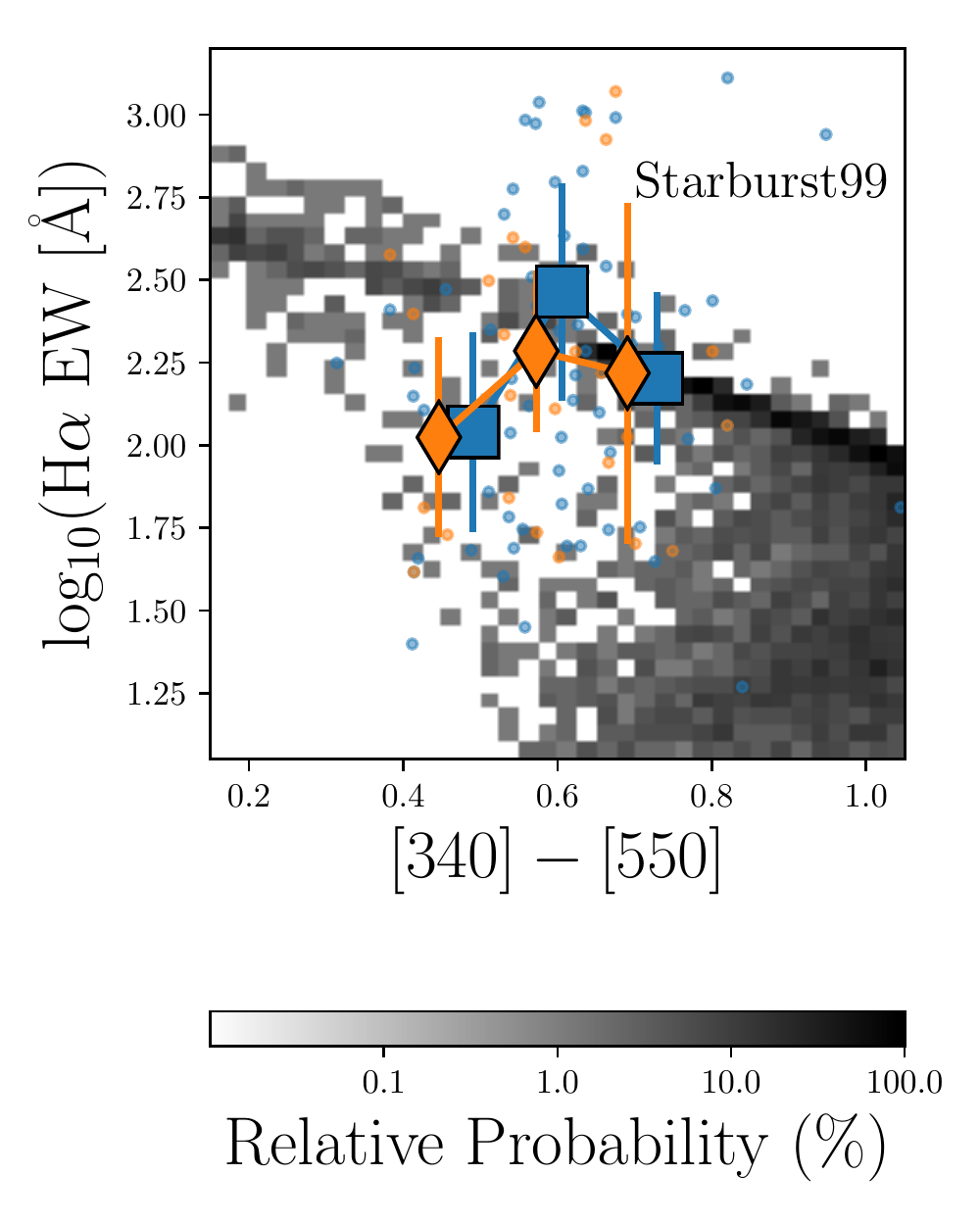}
\includegraphics[trim = 0 0 0 0, clip, scale=0.6]{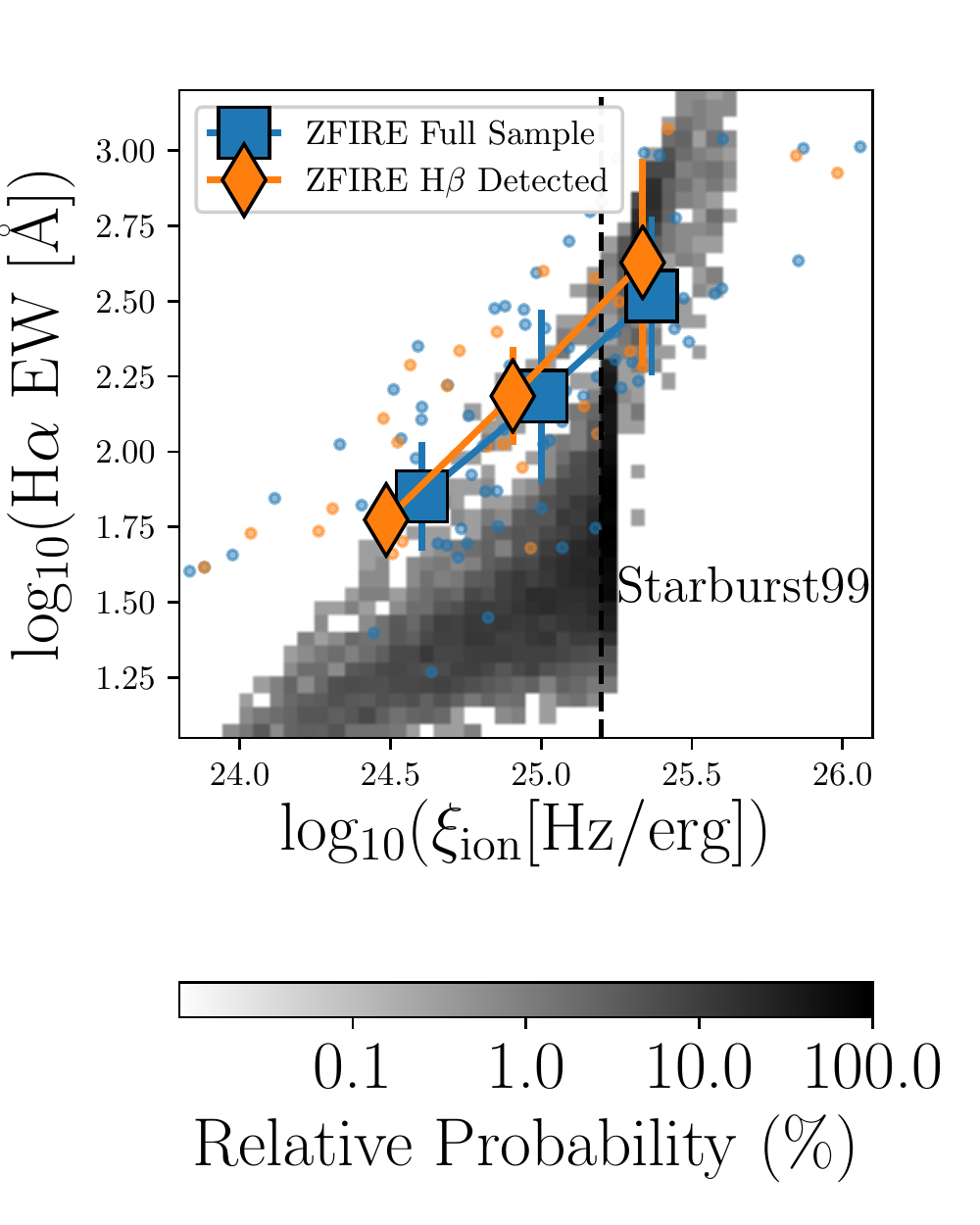}
\includegraphics[trim = 0 0 0 0, clip, scale=0.6]{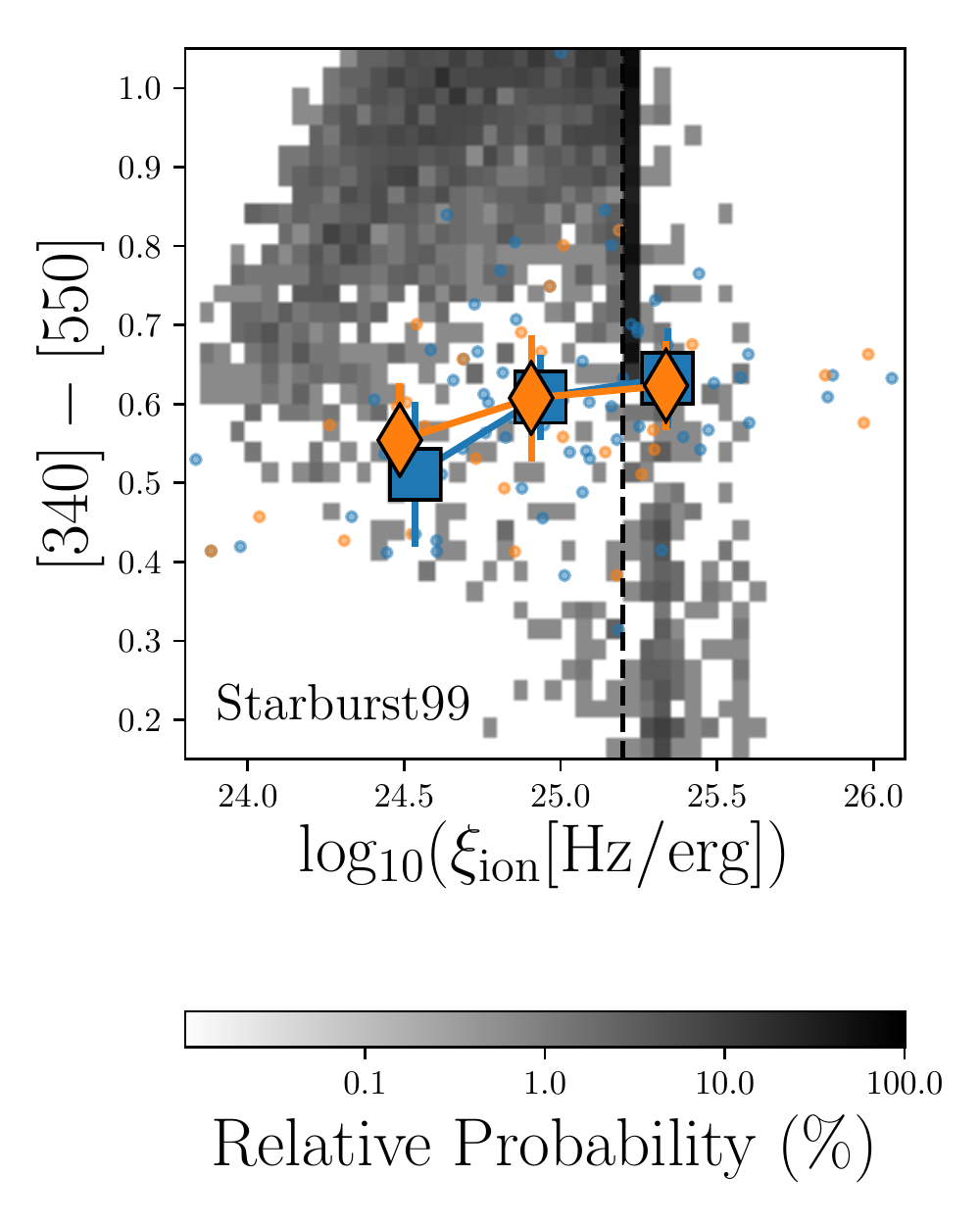}
\caption{ Probability distribution of 10,000 random realizations of 1000 Starburst99 model galaxies with constant+burst SFHs in {\bf left:}  \Halpha\ EW vs \boxfil\ color, {\bf center:}  \Halpha\ EW vs \xiion, and {\bf right:} \boxfil\ color vs \xiion\ space. Models are computed using Geneva stellar tracks at $\sim0.6$ \zsol\ that account for effects of stellar rotation with a \citet{Salpeter1955} like high-mass IMF. 
Multiple bursts (randomly chosen between $2-10$) with randomly chosen strengths ($\times5-100$) and lengths ($\Delta t= 10-100$ Myr) are overlaid on the constant SFH models at random times between $t=200-3000$ Myr. 
All models are normalized to the total mass formed by a 1\msol/yr constant SFH model at $z\sim2$ and are sampled in 1Myr time steps. 
The final grid contains $\sim1.6\times10^{6}$ steps between $1500-3100$ Myr, out of which 10,000 random iterations are selected with replacement. 2D density distribution of the selected values are shown as a relative probability distribution by the gray scale 2D histogram. 
\label{fig:ew_color_xiion_simulation}}
\end{figure*}

In Figure \ref{fig:ew_color_xiion_simulation}, we show the 2D density distribution of our randomly sampled iterations.  
Random time-sampling of constant+burst Starburst99 model galaxies is unable to reproduce the observed distribution of the $z\sim2$ galaxies in \xiion, \Halpha\ EW, and \boxfil\ space. 
In Figure \ref{fig:ew_color_xiion_models} we showed that model tracks of star-bursts do trace the observed distribution of galaxies in this space, however, given the very fast evolution of model tracks, not all values are equally likely. 
Thus, our random sampling exercise demonstrates that in a Universe where galaxies undergo bursts at random times, it is unlikely to preferentially observe galaxies with high \Halpha\ EWs, low \xiion, and blue optical colors. 

We also note that our simulations are quite simplistic and variations in Z, IMF, and other stellar model properties could lead to systematic limitations in our comparison of our simulations to the observed data. 
Implementing SED fitting of photometric data using non-parametric SFHs would allow us to probe the variation in SFH of individual galaxies and investigate under exactly what conditions of stellar properties we could reproduce the observables. We leave this to future work.

\section{Discussion}
\label{sec:discussion}

\subsection{Observed correlations of \xiion}
\label{sec:discussion_xiion_obs}

In Section \ref{sec:xi_ion_observed_correlations} we explored the variation of \xiion\ with various galaxy observables/properties. 
Our sample showed evidence for an enhancement of \xiion\ at $\beta<-1.5$, similar to previous observations \citep[e.g.,][]{Bouwens2016c,Shivaei2018} and reaches the canonical $log_{10}(\xi_{ion}[Hz/erg])\sim25.2$ value \citep{Robertson2013} at $\beta\sim2.0$. 
Thus, we expect an enhanced \xiion\ for galaxies with $\beta<-2.0$.  
Current observational constraints suggest $z>6$ galaxies to have bluer UV slopes compared to their low-$z$ counterparts \cite[e.g.,][]{Bouwens2009}, which may suggest an enhanced \xiion\ at $z>6$.

$\beta$ correlates with the UV/IR ratio and the UV reprocessed light in the far-IR making it a suitable tracer for dust attenuation \citep{Meurer1999}. 
At $z>2$, the infra-red to UV flux ratio is shown to correlate with $\beta$ and UV magnitude \citep{Bouwens2016b} and therefore it is possible for UV bright galaxies to have higher $\beta$ values. 
Therefore, is the enhancement of \xiion\ at lowest $\beta$ values a result of enhancement of \xiion\ at faint UV magnitudes?

The relationship between \xiion\ and UV magnitude is important to constrain the processes that dominated the ionizing photon budget in the $z>6$ Universe.
Faint UV galaxies are generally expected to have driven the reionization of the Universe due to their high number density and high Lyman-continuum leakage \citep[e.g.,][]{Duncan2015}, however, recent empirically motivated models suggest that massive UV bright galaxies contributed to the bulk of the reionization budget \citep{Naidu2019}. 

Once reionization is collectively constrained using \xiion, UV luminosity density, and the Lyman continuum escape fraction, an evolution of \xiion\ with UV magnitude is currently not favored.
This has been verified by some studies which show that \xiion\ has no correlation with UV magnitude \citep[e.g.,][]{Bouwens2016c,Shivaei2018,Lam2019a}, thus it seems unlikely that faint UV sources provide an additional contribution to reionization through elevated production of ionizing photons compared to UV bright sources.  
Our 80\% completeness in UV magnitude at $-18.8$ is brighter than observed $z\sim6$ median UV magnitude of $\mathrm{M_{UV}}\sim-17.5$ \citep{Bouwens2017}, therefore, galaxies fainter than our detection level are required to link with UV magnitude vs \xiion\ trends of galaxies observed in the reionization epoch.

An enhancement of \xiion\ at low stellar masses will also have implications to reionization processes in the $z>6$ Universe. 
At $z\sim2$, our lowest stellar mass bin shows an enhancement of \xiion, however, is also at the $\sim80\%$ mass completeness level of our survey. 
Therefore, similar to other $z\sim2$ studies \citep{Matthee2017,Shivaei2018} we cannot provide any constraints on whether there is an enhancement of \xiion\ at $\log_{10}(M_*/M\odot)\lesssim9.0$.

Accurate mass estimates require rest-frame optical coverage with $\lambda\gtrsim5000\AA$ \citep{Conroy2013}, thus $z>4$ stellar mass estimates derived purely from \emph{HST} photometry may lead to biases. 
Therefore deep \emph{Spitzer} or future \emph{JWST} observations of low mass star-forming galaxies at $z>4$ are crucial to determine, whether if there is a systematic increase in \xiion\ at lower masses leading up to the reionization era of the Universe.

Our set B galaxies showed evidence for a moderate positive correlation of \xiion\ with \OIII/\Halpha\ ratio. 
This correlation is driven by the presence of a strong correlation between \OIII\ and \Halpha\ fluxes with $r_s, p_s=0.9,2.6\times10^{-20}$ (which is weaker for set A galaxies; $r_s, p_s=0.7,6.7\times10^{-9}$) and the absence of a statistically significant correlation between \OIII\ flux and UV luminosity ($r_s, p_s=-0.3,0.06$). Set A galaxies show a moderate statistically significant negative correlation between \OIII\ flux and UV luminosity ($r_s, p_s=-0.4,5\times10^{-4}$).

\citet{Shivaei2018} demonstrated an enhancement of \xiion\ for galaxies with high [\hbox{{\rm O}\kern 0.1em{\sc iii}}]/\OII\ ratios, high [\hbox{{\rm O}\kern 0.1em{\sc iii}}]/\Hbeta\ ratios, and low \NII/\Halpha\ ratios. 
Similarly, \citet{Tang2018} showed \xiion\ to positively correlate with \OIII\ EW, and for \OIII\ EW to positively correlate with [\hbox{{\rm O}\kern 0.1em{\sc iii}}]/\OII. 
This translates to galaxies with high ionization parameter and/or low stellar metallicity having high \xiion. 
If galaxies have higher hydrogen ionizing photon densities compared to their hydrogen densities, at fixed SFR and ISM conditions naturally the ionizing photon production rate would be higher. 
Therefore, such an enhancement of \xiion\ at higher ionization parameter is expected and is possibly driven by the harder ionizing spectrum generated by the low metallicity stars due to less metal blanketing in stellar atmospheres and conservation of angular momentum due to weaker optically thick stellar winds leading to longer main-sequence lifetimes \citep[e.g.,][]{Eldridge2017}.

Our analysis did not show strong evidence for \xiion\ to vary as a function of UV+IR SFR. 
Since both SFR and \xiion\ are sensitive to the production rate of ionizing photons, a correlation between \Halpha\ SFR and \xiion\ is expected. 
However, \xiion\ is also sensitive to the stellar mass of young stars which contribute to the UV luminosity, and thus is a proxy for the sSFR. 
In the stellar-mass star-formation relation \citep{Tomczak2016}, high mass galaxies show high SFRs, therefore, it is reasonable to expect \xiion\ to also show a flat distribution with the SFR.
In terms of time evolution of \xiion, stellar population models with parametric SFHs follow a smooth evolution and in a constant SFH scenario \xiion\ will stabilize once the UV luminosity stabilizes. 
If galaxies undergo sudden bursts in their SFHs, the increase in SFR will be followed by an immediate increase in \xiion\ for a short period of time, after which \xiion\ will reduce and stabilize independent of the SFR.


\subsection{The completeness of our observed sample}
\label{sec:discussion_xiion_completeness}

In this analysis we presented 130 \Halpha\ emitters selected for spectroscopy from the ZFOURGE survey with a 80\% stellar  mass completeness at $log(M_{*}/M_{\odot})\sim9.3$. 
Our MOSFIRE spectroscopy sampled the $z\sim2$ large scale structure in the COSMOS field \citep{Spitler2012,Yuan2014} and our \Halpha\ spectroscopic detection rate is similar to within 1\% \citep{Nanayakkara2016} of the rate expected by the photometric redshift probability distribution functions computed using EAZY \citep{Brammer2008}. 
Additionally, we found that the \Halpha\ S/N of our sample peak at $\sim20$ and that $\sim81\%$ of our sample show a S/N of $>10$ reaching a $3\sigma$ \Halpha\ SFR detection limit at $\sim4$\msol/yr. 
Thus, we conclude that our sample has a high spectroscopic completeness based on our stellar mass/$Ks$ magnitude based photometric pre-selection.

\begin{figure}
\includegraphics[trim = 0 0 0 0, clip, scale=0.6]{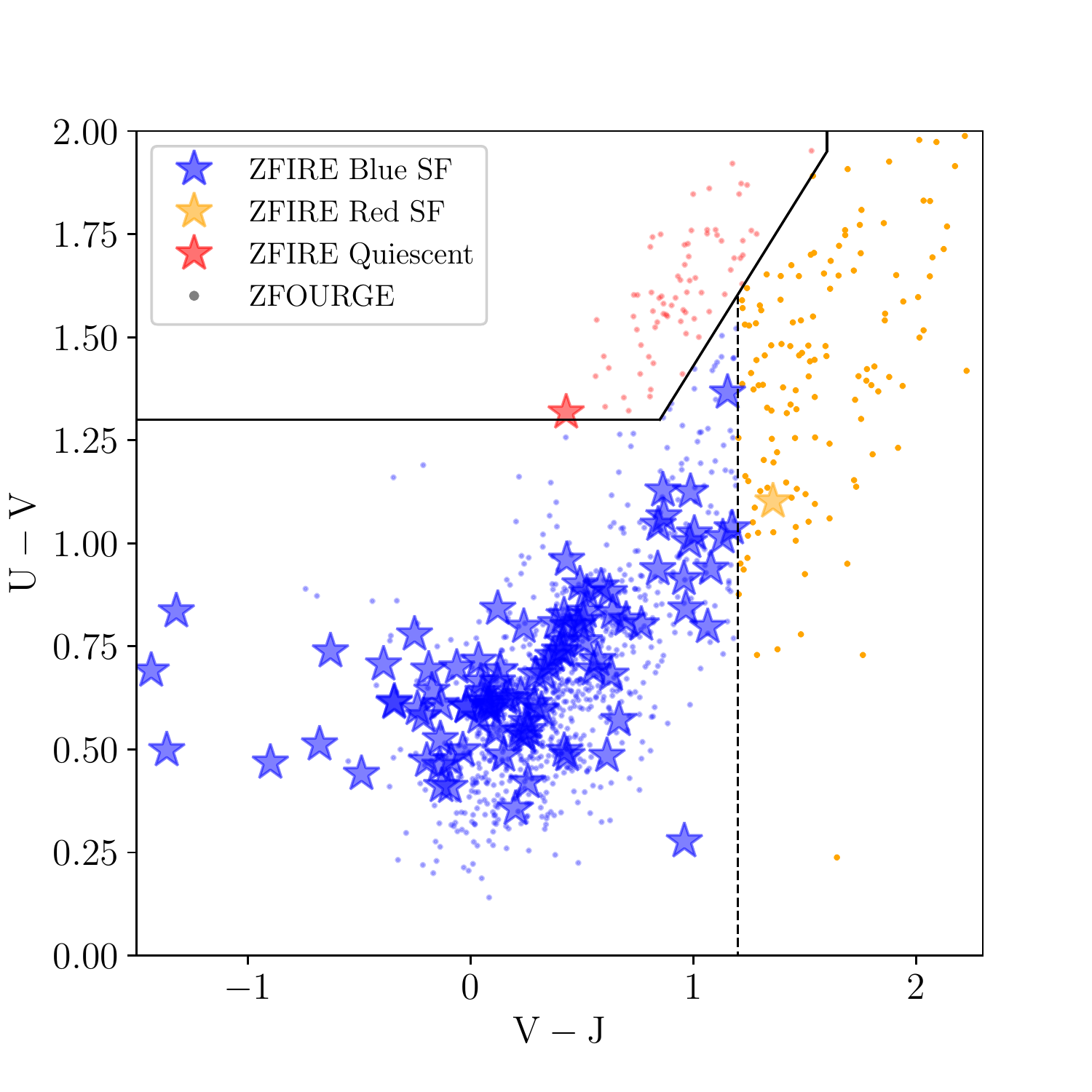}
\caption{ The rest-frame $U-V$ vs $V-J$ color distribution of the 102 ZFIRE $z\sim2$ galaxies used in the \xiion, \Halpha\ EW, and \boxfil\ color analysis. We also show the ZFOURGE parent population at $1.90<z_{EAZY}<2.66$ with $\log_{10}(M_*/M_\odot)>9.3$ and $Ks<24.78$. 
Galaxies are divided into blue star-forming (SF), red star-forming, and quiescent bins following \citet{Spitler2014} criteria. All rest-frame colors are derived using EAZY \citep{Brammer2008}. Our observed sample show a strong bias towards blue star-forming galaxies in this epoch. 
\label{fig:UVJ}}
\end{figure}

In Figure \ref{fig:UVJ}, we show the EAZY derived rest-frame $U-V$ vs $V-J$ colors of our spectroscopic sample with \Halpha\ EW measurements. 
Compared to the ZFOURGE COSMOS sample at $1.90<z_{EAZY}<2.66$ (the redshift window where \Halpha\ falls in MOSFIRE $K$ band) with $\log_{10}(M_*/M_\odot)>9.3$ and $Ks<24.78$ (which are respectively the $\sim80\%$ stellar-mass and $Ks$ completeness of the spectroscopic sample), the majority of galaxies used in this analysis are blue star-forming systems.
The $U-V$ vs $V-J$ color space can be used to distinguish between red (dusty) star-forming galaxies and passively evolving galaxies \citep[e.g.,][]{Labbe2007,Williams2009,Spitler2014}. 
Our \Halpha\ EW lacks red star-forming systems, which at high-$z$ tend to be high-mass or high-SFR galaxies 
\citep{Straatman2016}. The bluest $V-J$ colors of our sample are dominated by the lowest mass systems.

The lack of red star-forming galaxies in our \Halpha\ EW sample may translate to a lack of galaxies with low sSFRs. 
Low sSFR galaxies would have low \Halpha\ EWs, low \xiion, and redder \boxfil\ colors. 
Therefore, including red star-forming galaxies in our sample may move the average trends towards regions populated by exponentially declining SFHs with low $\tau$ values. 
However, our parametric SFHs or burst SFH simulation results will still not agree with the individual nor average trends of blue star-forming galaxies. 
Within the context of blue star-forming galaxies at $z\sim2$, we can rule out selection effects to have a strong influence on our observed correlations of \xiion\ with various galaxy properties explored in this analysis.

\subsection{Dust related uncertainties in the \xiion\ distribution}
\label{sec:discussion_dust_uncertainties}

In Figure \ref{fig:xiion_distribution_z} we showed that $\sim84\%$ of our sample fall below $log_{10}(\xi_{ion}[Hz/erg])=25.2$, which is estimated as the \xiion\ required to reionize the Universe by $z\sim6$ under current observational constraints \citep{Robertson2013}. 
Our distribution of \xiion\ is also similar to the analysis by \citet{Shivaei2018}, which used MOSFIRE spectroscopic data from the MOSDEF survey \citep{Kriek2015}. 
In terms of spectroscopic completeness and accurate \Halpha\ line flux estimates through high-quality spectroscopy and dust corrections, both our and \citet{Shivaei2018} analyses provide strong constraints to the \xiion\ in the $z\sim2$ Universe for stellar mass complete samples and also agrees well with narrow band emission line selected analysis by \citet{Matthee2017b}.

The choice of the dust attenuation law plays a role on \xiion\ measurements. If the UV luminosity is corrected using an attenuation curve steeper than that of \citet{Calzetti2000}, the dust corrected UV luminosity will be lower, which will increase the observed \xiion\ \cite[e.g.,][]{Bouwens2016b}. 
In terms of uncertainties related to the choice of the dust attenuation curve, \citet{Shivaei2018} demonstrated that effects to \xiion\ to be in the order of $\lesssim0.2$ dex. 
Secondary dependence of dust attenuation on metallicity of galaxies \citep[e.g.,][]{Reddy2018} may introduce additional complexities to the choice of the dust law, thus, with selective selection of dust attenuation law based on galaxy parameters may contribute to higher intrinsic \xiion.

Differential attenuation between nebular and stellar components also contribute to additional uncertainties in \xiion\ measurements. 
If the young stars reside near the stellar birth clouds, the nebular component in galaxies will undergo extra attenuation compared to the older stellar regions that contribute to the stellar continuum \citep[e.g.,][]{Calzetti1994}. 
Accurate determination of this absorption factor requires multiple Balmer emission line ratios, and may also show a dependence on galaxy properties such as the SFR \citep{Reddy2015}. 


\subsection{$z$ evolution of \xiion}
\label{sec:discussion_z_evolution}


Our observed $z\sim2$ \xiion\ measurements are $\sim0.5$ dex smaller compared to $z>4$ estimates. 
\citet{Matthee2017b} argues that one possibility for this observed discrepancy could be a redshift evolution of \xiion. 
Such an evolution may be justified if the SFHs of galaxies at $z>4$ are either dominated by exponentially rising SFHs or if they are very chaotic with frequent star-bursts. 
As we discussed in Section \ref{sec:star_bursts}, starbursts would drive \xiion\ to increase rapidly within shorter time-scales driven by an increase in the number of hydrogen ionizing photons. 
Thus, in this scenario we would expect $z\sim2$ star-forming galaxies to have exponentially declining SFHs dominated by relatively older stellar populations with high-UV luminosity, which would decrease \xiion.

In Section \ref{sec:star_bursts}, we discussed in detail on how star-bursts affect the evolution of \xiion. 
Even with single or multiple star-bursts with varying strengths, the time window in which \xiion\ would reach $log_{10}(\xi_{ion}[Hz/erg])>25.5$ is very short and is within time-scales of a few Myr. 
This is driven by the rapidly increasing contribution from the O and B type stars to the UV luminosity, which takes a longer time to stabilize compared to the more massive O stars that contribute to the hydrogen ionizing photons.   
Therefore, even if multiple star-bursts do contribute to an increase of \xiion\ in $z>4$ galaxies, the effects will be relatively short lived and it is unlikely for star-bursts to drive the high \xiion\ measurements.

\subsubsection{Selection effects in high-$z$ \xiion\ estimates}
\label{sec:discussion_high_z_selection_effects}

Driven by expectations from current stellar population models, we expect differences in selection functions of galaxies to play a dominant role in driving the differences in \xiion\ between $z\sim2$ and $z>4$. 
Observational estimates of \xiion\ at $z\sim7$ are obtained through highly selective samples of extreme \OIII+\Hbeta\ emitters \citep{Stark2015c}, thus the observed high \xiion\ may not represent typical galaxy populations in the reionization era.    

$z\gtrsim4$ photometric samples selected based on color selection through strong \Halpha+\NII\ contamination on the Spitzer/IRAC bands \citep[e.g.,][]{Shim2011,Bouwens2016c} would be biased towards strong \Halpha\ emitters. 
Additionally, the IRAC bands used to estimate the \Halpha\ flux is contaminated by \NII\ and \SII\ emission lines. 
Hard ionizing radiation fields and high ISM pressures in young stellar systems may lead to enhancements in \NII/\Halpha\ ratios \citep[e.g.,][]{Kewley2016}, which could lead to overestimate the \Halpha\ flux if a fixed nonevolving \NII/\Halpha\ ratio is used for the correction.

\citet{Lam2019a} spectroscopic sample is based on MUSE Deep \citep{Inami2017} and MUSE Wide \citep{Urrutia2019} surveys, where spectroscopic redshifts of galaxies at $z>4$ are primarily determined through Lyman-$\alpha$.    
\citet{Nakajima2016} sample also consists of Lyman-$\alpha$ emitters and Lyman break galaxies, thus introduce a strong sample selection bias \citep[e.g][]{Erb2016}.

$z\sim2$ galaxies based on extreme \OIII\ emitters show \xiion\ typical of $z>4$ samples and strong positive correlations with \OIII\ EW \citep{Tang2018}. 
Therefore it is likely that current high-$z$ \xiion\ measurements are biased towards strong line emitters.   
Additionally, dust attenuation uncertainties further complicates \xiion\ estimates at these redshifts since such properties for most of these galaxies at $z>3$ are not well constrained.

The observed correlation of \xiion\ with \OIII, and \Halpha\ EWs \citep[see also][]{Tang2018} suggest that current $z>4$ \xiion\ measurements may be biased towards strong line emitters. 
If a majority of galaxies at higher redshifts do show strong ionizing properties, the observed high \xiion\ of $z>4$ galaxies may be typical of the high-$z$ Universe. 
However, recent results have demonstrated that the observed diversity of galaxies in the $z\gtrsim4$ Universe is higher than what was previously expected \citep[e.g.,][]{Spitler2014,Straatman2014,Glazebrook2017,Schreiber2018b,Schreiber2018,Wang2019}. 
Therefore, deeper spectroscopic explorations of the $z>4$ Universe is essential to build up representative samples of galaxies to accurately determine if there is an enhancement of \xiion\ with $z$.

\subsubsection{Expectation from current stellar population models}
\label{sec:discussion_high_z_stellar_pop_effects}

In addition to selection effects biasing $z>4$ observations, it is also important to consider if current stellar population models lack sufficient amount of ionizing photons to reproduce high \xiion. 
Models may lack mechanisms/stellar types that may be prominent in galaxies in the early Universe that contribute to an increase of hydrogen ionizing photons. 
$z\sim0$ \citep[e.g.,][]{Kewley2001,Senchyna2017} and $z\sim2-4$ \citep[e.g.,][]{Nanayakkara2019} studies have shown that stellar population models may lack mechanisms that produce high energy photons in the EUV, which are required to produce observed emission line ratios such as \SII/\Halpha\ and observed \HeII\ spectral features (also see Section 5.3 of \citealt{Kewley2019} for a detailed discussion on current limitations of stellar populations models). 
Including effects of X-ray binaries \citep[e.g.,][]{Schaerer2019} and stripped stars \citep[e.g.,][]{Gotberg2019a} in stellar population synthesis models have shown to increase the production of ionizing photons.
Therefore, self consistent treatment of stellar evolution with rotation and binaries that contribute to such phenomena are crucial to make strong constraints on the nature of stellar populations in high-$z$ galaxies.  

Additionally, shallower slopes at the high mass end of the stellar IMF in galaxies in the early Universe will contribute to extra ionizing photons resulting in higher \xiion. 
However, currently there are no observational constraints to the high-mass stellar IMF at higher redshifts and investigating changes in the IMF slope is beyond the scope of this work.

Given galaxies in the early Universe are likely dominated by low-metallicity stars, it is plausible for \xiion\ to systematically increase with redshift.  
Within the context of constant SFHs, sub-solar metallicity BPASS models show a higher \xiion\ at fixed metallicity compared to Starburst99 models. 
Constant SFH models with binaries from BPASSv2.2.1 show a strong dependence on stellar metallicity with the lowest stellar metallicity models showing the highest \xiion. 
At $t\sim1$ Gyr, $Z\leq1/5$ \zsol\ models show $log_{10}(\xi_{ion}[Hz/erg])\sim25.3-25.4$. 
Higher metallicity BPASS models do not consider effects of quasi homogeneous evolution \citep{Eldridge2017} and therefore show $log_{10}(\xi_{ion}[Hz/erg])\sim25.2$.

Starburst99 models used in our analysis showed an opposite effect, where $Z=0.002$ models show lower \xiion\ compared to $Z=0.014$ models (note that in Starburst99 \zsol=0.014 in contrast to \zsol=0.02 in BPASS models). 
The highest \xiion\ achieved by Starburst99 Geneva rotational \zsol\ stellar tracks at $t=1$ Gyr with a constant SFH is $log_{10}(\xi_{ion}[Hz/erg])\sim25.2$. 
At higher metallicities, stars will loose angular momentum faster due to optically think winds, therefore, lower metallicity stars would have higher temperatures and longer main-sequence life-times \citep{Leitherer2014}.
However, this increase in ionizing photons at lower metallicities is counteracted by the high abundance of W-R stars at \zsol.
Therefore, the increase in \xiion\ with $Z$ in Starburst99 models is modest ($\sim0.07$ dex increase between 1/7th \zsol\ to \zsol\ models). 
In BPASS models, effects of mass transfer between close binary stars results outer layers of massive red super-giants to be removed efficiently leading to a higher fraction of W–R stars and/or low-mass helium stars. 
Therefore, even at lower metallicities there is an abundance of W-R stars in BPASS models.


\section{Conclusions}
\label{sec:conclusion}

In this analysis, we presented an analysis of the ionizing photon production efficiency (\xiion) of a mass/$Ks$ magnitude selected sample of star-forming galaxies at $z\sim2$ observed by the ZFIRE survey using KECK/MOSFIRE. 
We analyzed how the \xiion\ correlates with observed/derived galaxy properties and combined our analysis of \xiion\ with \Halpha\ EW and \boxfil\ colors, a commonly used diagnostic to analyze the stellar populations properties of star-forming galaxies \cite[e.g.,][]{Nanayakkara2017}.  

Our main conclusions are as follows:
\begin{itemize}
	\item The distribution of \xiion\ of our sample is similar to similar studies at $z\sim2$, with a majority of galaxies falling below the canonical $log_{10}(\xi_{ion}[Hz/erg])=25.2$ required to reionize the Universe by $z=6$. 

	\item By analyzing the \xiion\ correlation of our sample with galaxy properties such as UV continuum slope $\beta$, UV magnitude, stellar mass, \OIII/\Halpha\ ratio, and UV+IR and \Halpha\ SFRs, we demonstrated that our results agree well with other studies of star-forming galaxies at $z\sim2$.

	\item We combined our analysis of \xiion\ with \Halpha\ EW and rest-frame optical colors and analyzed the distribution of our sample with smooth SFH predictions from BPASSv2.2.1 stellar population models. We found that stellar models cannot self-consistently predict the observed the distribution of galaxies in \xiion, \Halpha\ EW, and \boxfil\ color space. At fixed \xiion\ the models always show lower \Halpha\ EWs and redder \boxfil\ colors compared to the data. 

	\item We used Starburst99 stellar population models with various star-burst properties to perform Monte-Carlo simulations of galaxies in \xiion, \Halpha\ EW, and \boxfil\ color space. Our random sampling of galaxies showed that statistically it was unlikely to randomly select galaxies that populate our observed distribution.


\end{itemize}

Our analysis demonstrated that, within the context of the simple SFHs explored by us, the stellar population models cannot self-consistently predict the observed the distribution of $z\sim2$ galaxies in \xiion, \Halpha\ EW, and \boxfil\ color space. 
This may translate to a lack of hydrogen ionizing photons in UV bright galaxies in stellar population models. Thus stellar population models may require additional changes to increase the ionizing photon output. 
In future we will extend the analysis presented here using {\tt Prospector} \citep{Leja2017} to investigate the individual SFHs of our sample at $z\sim2$ and determine under which conditions the observed distribution of galaxies could be reproduced by non-parametric SFHs.

\acknowledgements{The authors wish to recognize and acknowledge the very significant cultural role and reverence that the summit of Maunakea has always had within the indigenous Hawaiian community.
We thank the anonymous referee for taking their valuable time to provide thorough and constructive comments.
We thank Irene Shivaei for providing the \xiion\ measurements used in the \citet{Shivaei2018} analysis. 
TN, JB, and RB acknowledges the Nederlandse Organisatie voor Wetenschappelijk Onderzoek (NWO) top grant TOP1.16.057. 
JB acknowledges support by Fundação para a Ciência e a Tecnologia (FCT) through national funds (UID/FIS/04434/2013) and Investigador FCT contract IF/01654/2014/CP1215/CT0003, and by FEDER through COMPETE2020 (POCI-01-0145-FEDER-007672).
GGK acknowledges the support of the Australian Research Council through the Discovery Project DP170103470. 
Parts of this research were supported by the Australian Research Council Centre of Excellence for All Sky Astrophysics in 3 Dimensions (ASTRO 3D), through project number CE170100013.}

Facilities: \facility{Keck:I (MOSFIRE)}

\bibliographystyle{apj}
\bibliography{zfire_xiion_bibliography.bib}

\begin{thebibliography}{}
\expandafter\ifx\csname natexlab\endcsname\relax\def\natexlab#1{#1}\fi

\bibitem[{{Alcorn} {et~al.}(2016){Alcorn}, {Tran}, {Kacprzak}, {Nanayakkara},
  {Straatman}, {Yuan}, {Allen}, {Cowley}, {Dav{\'e}}, {Glazebrook}, {Kewley},
  {Labb{\'e}}, {Quadri}, {Spitler}, \& {Tomczak}}]{Alcorn2016}
{Alcorn}, L.~Y., {Tran}, K.-V.~H., {Kacprzak}, G.~G., {et~al.} 2016, \apjl,
  825, L2

\bibitem[{{Barkana} \& {Loeb}(2006)}]{Barkana2006}
{Barkana}, R., \& {Loeb}, A. 2006, \mnras, 371, 395

\bibitem[{Beckwith {et~al.}(2006)Beckwith, Stiavelli, Koekemoer, Caldwell,
  Ferguson, Hook, Lucas, Bergeron, Corbin, Jogee, Panagia, Robberto, Royle,
  Somerville, \& Sosey}]{Beckwith2006}
Beckwith, S. V.~W., Stiavelli, M., Koekemoer, A.~M., {et~al.} 2006, {AJ}, 132,
  1729

\bibitem[{{Bouwens} {et~al.}(2015{\natexlab{a}}){Bouwens}, {Illingworth},
  {Oesch}, {Caruana}, {Holwerda}, {Smit}, \& {Wilkins}}]{Bouwens2015b}
{Bouwens}, R.~J., {Illingworth}, G.~D., {Oesch}, P.~A., {et~al.}
  2015{\natexlab{a}}, \apj, 811, 140

\bibitem[{{Bouwens} {et~al.}(2017){Bouwens}, {Oesch}, {Illingworth}, {Ellis},
  \& {Stefanon}}]{Bouwens2017}
{Bouwens}, R.~J., {Oesch}, P.~A., {Illingworth}, G.~D., {Ellis}, R.~S., \&
  {Stefanon}, M. 2017, \apj, 843, 129

\bibitem[{{Bouwens} {et~al.}(2016{\natexlab{a}}){Bouwens}, {Smit}, {Labb{\'e}},
  {Franx}, {Caruana}, {Oesch}, {Stefanon}, \& {Rasappu}}]{Bouwens2016c}
{Bouwens}, R.~J., {Smit}, R., {Labb{\'e}}, I., {et~al.} 2016{\natexlab{a}},
  \apj, 831, 176

\bibitem[{{Bouwens} {et~al.}(2009){Bouwens}, {Illingworth}, {Franx}, {Chary},
  {Meurer}, {Conselice}, {Ford}, {Giavalisco}, \& {van Dokkum}}]{Bouwens2009}
{Bouwens}, R.~J., {Illingworth}, G.~D., {Franx}, M., {et~al.} 2009, \apj, 705,
  936

\bibitem[{{Bouwens} {et~al.}(2015{\natexlab{b}}){Bouwens}, {Illingworth},
  {Oesch}, {Trenti}, {Labb{\'e}}, {Bradley}, {Carollo}, {van Dokkum},
  {Gonzalez}, {Holwerda}, {Franx}, {Spitler}, {Smit}, \& {Magee}}]{Bouwens2015}
{Bouwens}, R.~J., {Illingworth}, G.~D., {Oesch}, P.~A., {et~al.}
  2015{\natexlab{b}}, \apj, 803, 34

\bibitem[{{Bouwens} {et~al.}(2016{\natexlab{b}}){Bouwens}, {Aravena},
  {Decarli}, {Walter}, {da Cunha}, {Labb{\'e}}, {Bauer}, {Bertoldi}, {Carilli},
  {Chapman}, {Daddi}, {Hodge}, {Ivison}, {Karim}, {Le Fevre}, {Magnelli},
  {Ota}, {Riechers}, {Smail}, {van der Werf}, {Weiss}, {Cox}, {Elbaz},
  {Gonzalez-Lopez}, {Infante}, {Oesch}, {Wagg}, \& {Wilkins}}]{Bouwens2016b}
{Bouwens}, R.~J., {Aravena}, M., {Decarli}, R., {et~al.} 2016{\natexlab{b}},
  \apj, 833, 72

\bibitem[{{Brammer} {et~al.}(2008){Brammer}, {van Dokkum}, \&
  {Coppi}}]{Brammer2008}
{Brammer}, G.~B., {van Dokkum}, P.~G., \& {Coppi}, P. 2008, \apj, 686, 1503

\bibitem[{{Bruzual} \& {Charlot}(2003)}]{Bruzual2003}
{Bruzual}, G., \& {Charlot}, S. 2003, \mnras, 344, 1000

\bibitem[{{Calzetti} {et~al.}(2000){Calzetti}, {Armus}, {Bohlin}, {Kinney},
  {Koornneef}, \& {Storchi-Bergmann}}]{Calzetti2000}
{Calzetti}, D., {Armus}, L., {Bohlin}, R.~C., {et~al.} 2000, \apj, 533, 682

\bibitem[{{Calzetti} {et~al.}(1994){Calzetti}, {Kinney}, \&
  {Storchi-Bergmann}}]{Calzetti1994}
{Calzetti}, D., {Kinney}, A.~L., \& {Storchi-Bergmann}, T. 1994, \apj, 429, 582

\bibitem[{{Cardelli} {et~al.}(1989){Cardelli}, {Clayton}, \&
  {Mathis}}]{Cardelli1989}
{Cardelli}, J.~A., {Clayton}, G.~C., \& {Mathis}, J.~S. 1989, \apj, 345, 245

\bibitem[{Chabrier(2003)}]{Chabrier2003}
Chabrier, G. 2003, Publications of the Astronomical Society of the Pacific,
  115, pp. 763

\bibitem[{{Conroy}(2013)}]{Conroy2013}
{Conroy}, C. 2013, \araa, 51, 393

\bibitem[{{Cowley} {et~al.}(2016){Cowley}, {Spitler}, {Tran}, {Rees},
  {Labb{\'e}}, {Allen}, {Brammer}, {Glazebrook}, {Hopkins}, {Juneau},
  {Kacprzak}, {Mullaney}, {Nanayakkara}, {Papovich}, {Quadri}, {Straatman},
  {Tomczak}, \& {van Dokkum}}]{Cowley2016}
{Cowley}, M.~J., {Spitler}, L.~R., {Tran}, K.-V.~H., {et~al.} 2016, \mnras,
  457, 629

\bibitem[{{Draine}(2011)}]{Draine2011}
{Draine}, B.~T. 2011, {Physics of the Interstellar and Intergalactic Medium}

\bibitem[{{Duncan} \& {Conselice}(2015)}]{Duncan2015}
{Duncan}, K., \& {Conselice}, C.~J. 2015, \mnras, 451, 2030

\bibitem[{{Eldridge} {et~al.}(2017){Eldridge}, {Stanway}, {Xiao}, {McClelland},
  {Taylor}, {Ng}, {Greis}, \& {Bray}}]{Eldridge2017}
{Eldridge}, J.~J., {Stanway}, E.~R., {Xiao}, L., {et~al.} 2017, PASA, 34, e058

\bibitem[{{Erb} {et~al.}(2016){Erb}, {Pettini}, {Steidel}, {Strom}, {Rudie},
  {Trainor}, {Shapley}, \& {Reddy}}]{Erb2016}
{Erb}, D.~K., {Pettini}, M., {Steidel}, C.~C., {et~al.} 2016, \apj, 830, 52

\bibitem[{{Finkelstein} {et~al.}(2015){Finkelstein}, {Ryan}, {Papovich},
  {Dickinson}, {Song}, {Somerville}, {Ferguson}, {Salmon}, {Giavalisco},
  {Koekemoer}, {Ashby}, {Behroozi}, {Castellano}, {Dunlop}, {Faber}, {Fazio},
  {Fontana}, {Grogin}, {Hathi}, {Jaacks}, {Kocevski}, {Livermore}, {McLure},
  {Merlin}, {Mobasher}, {Newman}, {Rafelski}, {Tilvi}, \&
  {Willner}}]{Finkelstein2015b}
{Finkelstein}, S.~L., {Ryan}, Russell~E., J., {Papovich}, C., {et~al.} 2015,
  \apj, 810, 71

\bibitem[{{Giacconi} {et~al.}(2001){Giacconi}, {Rosati}, {Tozzi}, {Nonino},
  {Hasinger}, {Norman}, {Bergeron}, {Borgani}, {Gilli}, {Gilmozzi}, \&
  {Zheng}}]{Giacconi2001}
{Giacconi}, R., {Rosati}, P., {Tozzi}, P., {et~al.} 2001, \apj, 551, 624

\bibitem[{{Glazebrook} {et~al.}(2017){Glazebrook}, {Schreiber}, {Labb{\'e}},
  {Nanayakkara}, {Kacprzak}, {Oesch}, {Papovich}, {Spitler}, {Straatman},
  {Tran}, \& {Yuan}}]{Glazebrook2017}
{Glazebrook}, K., {Schreiber}, C., {Labb{\'e}}, I., {et~al.} 2017, \nat, 544,
  71

\bibitem[{{G{\"o}tberg} {et~al.}(2019){G{\"o}tberg}, {de Mink}, {Groh},
  {Leitherer}, \& {Norman}}]{Gotberg2019a}
{G{\"o}tberg}, Y., {de Mink}, S.~E., {Groh}, J.~H., {Leitherer}, C., \&
  {Norman}, C. 2019, \aap, 629, A134

\bibitem[{{Gunawardhana} {et~al.}(2011){Gunawardhana}, {Hopkins}, {Sharp},
  {Brough}, {Taylor}, {Bland-Hawthorn}, {Maraston}, {Tuffs}, {Popescu},
  {Wijesinghe}, {Jones}, {Croom}, {Sadler}, {Wilkins}, {Driver}, {Liske},
  {Norberg}, {Baldry}, {Bamford}, {Loveday}, {Peacock}, {Robotham}, {Zucker},
  {Parker}, {Conselice}, {Cameron}, {Frenk}, {Hill}, {Kelvin}, {Kuijken},
  {Madore}, {Nichol}, {Parkinson}, {Pimbblet}, {Prescott}, {Sutherland},
  {Thomas}, \& {van Kampen}}]{Gunawardhana2011}
{Gunawardhana}, M.~L.~P., {Hopkins}, A.~M., {Sharp}, R.~G., {et~al.} 2011,
  \mnras, 415, 1647

\bibitem[{{Haydon} {et~al.}(2018){Haydon}, {Kruijssen}, {Hygate}, {Schruba},
  {Krumholz}, {Chevance}, \& {Longmore}}]{Haydon2018}
{Haydon}, D.~T., {Kruijssen}, J.~M.~D., {Hygate}, A. P.~S., {et~al.} 2018,
  arXiv e-prints, arXiv:1810.10897

\bibitem[{{Hoversten} \& {Glazebrook}(2008)}]{Hoversten2008}
{Hoversten}, E.~A., \& {Glazebrook}, K. 2008, \apj, 675, 163

\bibitem[{{Inami} {et~al.}(2017){Inami}, {Bacon}, {Brinchmann}, {Richard},
  {Contini}, {Conseil}, {Hamer}, {Akhlaghi}, {Bouch{\'e}}, {Cl{\'e}ment},
  {Desprez}, {Drake}, {Hashimoto}, {Leclercq}, {Maseda}, {Michel-Dansac},
  {Paalvast}, {Tresse}, {Ventou}, {Kollatschny}, {Boogaard}, {Finley},
  {Marino}, {Schaye}, \& {Wisotzki}}]{Inami2017}
{Inami}, H., {Bacon}, R., {Brinchmann}, J., {et~al.} 2017, \aap, 608, A2

\bibitem[{{Kacprzak} {et~al.}(2015){Kacprzak}, {Yuan}, {Nanayakkara},
  {Kobayashi}, {Tran}, {Kewley}, {Glazebrook}, {Spitler}, {Taylor}, {Cowley},
  {Labbe}, {Straatman}, \& {Tomczak}}]{Kacprzak2015}
{Kacprzak}, G.~G., {Yuan}, T., {Nanayakkara}, T., {et~al.} 2015, \apjl, 802,
  L26

\bibitem[{{Kennicutt}(1983)}]{Kennicutt1983}
{Kennicutt}, Jr., R.~C. 1983, \apj, 272, 54

\bibitem[{{Kewley} {et~al.}(2013){Kewley}, {Dopita}, {Leitherer}, {Dav{\'e}},
  {Yuan}, {Allen}, {Groves}, \& {Sutherland}}]{Kewley2013a}
{Kewley}, L.~J., {Dopita}, M.~A., {Leitherer}, C., {et~al.} 2013, \apj, 774,
  100

\bibitem[{{Kewley} {et~al.}(2001){Kewley}, {Dopita}, {Sutherland}, {Heisler},
  \& {Trevena}}]{Kewley2001}
{Kewley}, L.~J., {Dopita}, M.~A., {Sutherland}, R.~S., {Heisler}, C.~A., \&
  {Trevena}, J. 2001, \apj, 556, 121

\bibitem[{Kewley {et~al.}(2019)Kewley, Nicholls, \& Sutherland}]{Kewley2019}
Kewley, L.~J., Nicholls, D.~C., \& Sutherland, R.~S. 2019, Annual Review of
  Astronomy and Astrophysics, 57, null

\bibitem[{{Kewley} {et~al.}(2016){Kewley}, {Yuan}, {Nanayakkara}, {Kacprzak},
  {Tran}, {Glazebrook}, {Spitler}, {Cowley}, {Dopita}, {Straatman},
  {Labb{\'e}}, \& {Tomczak}}]{Kewley2016}
{Kewley}, L.~J., {Yuan}, T., {Nanayakkara}, T., {et~al.} 2016, \apj, 819, 100

\bibitem[{{Kriek} \& {Conroy}(2013)}]{Kriek2013}
{Kriek}, M., \& {Conroy}, C. 2013, \apjl, 775, L16

\bibitem[{{Kriek} {et~al.}(2015){Kriek}, {Shapley}, {Reddy}, {Siana}, {Coil},
  {Mobasher}, {Freeman}, {de Groot}, {Price}, {Sanders}, {Shivaei}, {Brammer},
  {Momcheva}, {Skelton}, {van Dokkum}, {Whitaker}, {Aird}, {Azadi}, {Kassis},
  {Bullock}, {Conroy}, {Dav{\'e}}, {Kere{\v s}}, \& {Krumholz}}]{Kriek2015}
{Kriek}, M., {Shapley}, A.~E., {Reddy}, N.~A., {et~al.} 2015, \apjs, 218, 15

\bibitem[{{Kuhlen} \& {Faucher-Gigu{\`e}re}(2012)}]{Kuhlen2012}
{Kuhlen}, M., \& {Faucher-Gigu{\`e}re}, C.-A. 2012, \mnras, 423, 862

\bibitem[{{Labb{\'e}} {et~al.}(2007){Labb{\'e}}, {Franx}, {Rudnick},
  {Schreiber}, {van Dokkum}, {Moorwood}, {Rix}, {R{\"o}ttgering}, {Trujillo},
  \& {van der Werf}}]{Labbe2007}
{Labb{\'e}}, I., {Franx}, M., {Rudnick}, G., {et~al.} 2007, The Astrophysical
  Journal, 665, 944

\bibitem[{{Lam} {et~al.}(2019){Lam}, {Bouwens}, {Labbe}, {Schaye}, {Schmidt},
  {Maseda}, {Bacon}, {Boogaard}, {Nanayakkara}, \& {Richard}}]{Lam2019a}
{Lam}, D., {Bouwens}, R.~J., {Labbe}, I., {et~al.} 2019, arXiv e-prints,
  arXiv:1902.02786

\bibitem[{{Leitherer} {et~al.}(2014){Leitherer}, {Ekstr{\"o}m}, {Meynet},
  {Schaerer}, {Agienko}, \& {Levesque}}]{Leitherer2014}
{Leitherer}, C., {Ekstr{\"o}m}, S., {Meynet}, G., {et~al.} 2014, \apjs, 212, 14

\bibitem[{{Leja} {et~al.}(2017){Leja}, {Johnson}, {Conroy}, {van Dokkum}, \&
  {Byler}}]{Leja2017}
{Leja}, J., {Johnson}, B.~D., {Conroy}, C., {van Dokkum}, P.~G., \& {Byler}, N.
  2017, \apj, 837, 170

\bibitem[{{M{\'a}rmol-Queralt{\'o}} {et~al.}(2016){M{\'a}rmol-Queralt{\'o}},
  {McLure}, {Cullen}, {Dunlop}, {Fontana}, \& {McLeod}}]{Marmol-Queralto2016}
{M{\'a}rmol-Queralt{\'o}}, E., {McLure}, R.~J., {Cullen}, F., {et~al.} 2016,
  \mnras, 460, 3587

\bibitem[{{Matthee} {et~al.}(2017{\natexlab{a}}){Matthee}, {Sobral}, {Best},
  {Khostovan}, {Oteo}, {Bouwens}, \& {R{\"o}ttgering}}]{Matthee2017b}
{Matthee}, J., {Sobral}, D., {Best}, P., {et~al.} 2017{\natexlab{a}}, \mnras,
  465, 3637

\bibitem[{{Matthee} {et~al.}(2017{\natexlab{b}}){Matthee}, {Sobral}, {Boone},
  {R{\"o}ttgering}, {Schaerer}, {Girard}, {Pallottini}, {Vallini}, {Ferrara},
  {Darvish}, \& {Mobasher}}]{Matthee2017}
{Matthee}, J., {Sobral}, D., {Boone}, F., {et~al.} 2017{\natexlab{b}}, \apj,
  851, 145

\bibitem[{McLean {et~al.}(2012)McLean, Steidel, Epps, Konidaris, Matthews,
  Adkins, Aliado, Brims, Canfield, Cromer, Fucik, Kulas, Mace, Magnone,
  Rodriguez, Rudie, Trainor, Wang, Weber, \& Weiss}]{McLean2012}
McLean, I.~S., Steidel, C.~C., Epps, H.~W., {et~al.} 2012, in Ground-based and
  Airborne Instrumentation for Astronomy {IV}, ed. I.~S. McLean, S.~K. Ramsay,
  \& H.~Takami, Vol. 8446 ({SPIE}-Intl Soc Optical Eng), 84460J--84460J--15

\bibitem[{{Messias} {et~al.}(2012){Messias}, {Afonso}, {Salvato}, {Mobasher},
  \& {Hopkins}}]{Messias2012}
{Messias}, H., {Afonso}, J., {Salvato}, M., {Mobasher}, B., \& {Hopkins}, A.~M.
  2012, \apj, 754, 120

\bibitem[{{Meurer} {et~al.}(1999){Meurer}, {Heckman}, \&
  {Calzetti}}]{Meurer1999}
{Meurer}, G.~R., {Heckman}, T.~M., \& {Calzetti}, D. 1999, \apj, 521, 64

\bibitem[{{Naidu} {et~al.}(2019){Naidu}, {Tacchella}, {Mason}, {Bose}, {Oesch},
  \& {Conroy}}]{Naidu2019}
{Naidu}, R.~P., {Tacchella}, S., {Mason}, C.~A., {et~al.} 2019, arXiv e-prints,
  arXiv:1907.13130

\bibitem[{{Nakajima} {et~al.}(2016){Nakajima}, {Ellis}, {Iwata}, {Inoue},
  {Kusakabe}, {Ouchi}, \& {Robertson}}]{Nakajima2016}
{Nakajima}, K., {Ellis}, R.~S., {Iwata}, I., {et~al.} 2016, \apj, 831, L9

\bibitem[{{Nakajima} {et~al.}(2017){Nakajima}, {Schaerer}, {Le Fevre},
  {Amorin}, {Talia}, {Lemaux}, {Tasca}, {Vanzella}, {Zamorani}, {Bardelli},
  {Grazian}, {Guaita}, {Hathi}, {Pentericci}, \& {Zucca}}]{Nakajima2017}
{Nakajima}, K., {Schaerer}, D., {Le Fevre}, O., {et~al.} 2017, ArXiv e-prints,
  arXiv:1709.03990

\bibitem[{{Nanayakkara} {et~al.}(2016){Nanayakkara}, {Glazebrook}, {Kacprzak},
  {Yuan}, {Tran}, {Spitler}, {Kewley}, {Straatman}, {Cowley}, {Fisher},
  {Labbe}, {Tomczak}, {Allen}, \& {Alcorn}}]{Nanayakkara2016}
{Nanayakkara}, T., {Glazebrook}, K., {Kacprzak}, G.~G., {et~al.} 2016, \apj,
  828, 21

\bibitem[{{Nanayakkara} {et~al.}(2017){Nanayakkara}, {Glazebrook}, {Kacprzak},
  {Yuan}, {Fisher}, {Tran}, {Kewley}, {Spitler}, {Alcorn}, {Cowley}, {Labbe},
  {Straatman}, \& {Tomczak}}]{Nanayakkara2017}
---. 2017, \mnras, 468, 3071

\bibitem[{{Nanayakkara} {et~al.}(2019){Nanayakkara}, {Brinchmann}, {Boogaard},
  {Bouwens}, {Cantalupo}, {Feltre}, {Kollatschny}, {Marino}, {Maseda},
  {Matthee}, {Paalvast}, {Richard}, \& {Verhamme}}]{Nanayakkara2019}
{Nanayakkara}, T., {Brinchmann}, J., {Boogaard}, L., {et~al.} 2019, \aap, 624,
  A89

\bibitem[{{Oke} \& {Gunn}(1983)}]{Oke1983}
{Oke}, J.~B., \& {Gunn}, J.~E. 1983, \apj, 266, 713

\bibitem[{{Persson} {et~al.}(2013){Persson}, {Murphy}, {Smee}, {Birk},
  {Monson}, {Uomoto}, {Koch}, {Shectman}, {Barkhouser}, {Orndorff}, {Hammond},
  {Harding}, {Scharfstein}, {Kelson}, {Marshall}, \& {McCarthy}}]{Persson2013}
{Persson}, S.~E., {Murphy}, D.~C., {Smee}, S., {et~al.} 2013, \pasp, 125, 654

\bibitem[{{Reddy} {et~al.}(2016){Reddy}, {Steidel}, {Pettini},
  {Bogosavljevi{\'c}}, \& {Shapley}}]{Reddy2016b}
{Reddy}, N.~A., {Steidel}, C.~C., {Pettini}, M., {Bogosavljevi{\'c}}, M., \&
  {Shapley}, A.~E. 2016, \apj, 828, 108

\bibitem[{{Reddy} {et~al.}(2015){Reddy}, {Kriek}, {Shapley}, {Freeman},
  {Siana}, {Coil}, {Mobasher}, {Price}, {Sanders}, \& {Shivaei}}]{Reddy2015}
{Reddy}, N.~A., {Kriek}, M., {Shapley}, A.~E., {et~al.} 2015, \apj, 806, 259

\bibitem[{{Reddy} {et~al.}(2018){Reddy}, {Oesch}, {Bouwens}, {Montes},
  {Illingworth}, {Steidel}, {van Dokkum}, {Atek}, {Carollo}, {Cibinel},
  {Holden}, {Labb{\'e}}, {Magee}, {Morselli}, {Nelson}, \&
  {Wilkins}}]{Reddy2018}
{Reddy}, N.~A., {Oesch}, P.~A., {Bouwens}, R.~J., {et~al.} 2018, \apj, 853, 56

\bibitem[{{Rees} {et~al.}(2016){Rees}, {Spitler}, {Norris}, {Cowley},
  {Papovich}, {Glazebrook}, {Quadri}, {Straatman}, {Allen}, {Kacprzak},
  {Labbe}, {Nanayakkara}, {Tomczak}, \& {Tran}}]{Rees2016}
{Rees}, G.~A., {Spitler}, L.~R., {Norris}, R.~P., {et~al.} 2016, \mnras, 455,
  2731

\bibitem[{{Robertson} {et~al.}(2015){Robertson}, {Ellis}, {Furlanetto}, \&
  {Dunlop}}]{Robertson2015}
{Robertson}, B.~E., {Ellis}, R.~S., {Furlanetto}, S.~R., \& {Dunlop}, J.~S.
  2015, \apj, 802, L19

\bibitem[{{Robertson} {et~al.}(2013){Robertson}, {Furlanetto}, {Schneider},
  {Charlot}, {Ellis}, {Stark}, {McLure}, {Dunlop}, {Koekemoer}, {Schenker},
  {Ouchi}, {Ono}, {Curtis-Lake}, {Rogers}, {Bowler}, \&
  {Cirasuolo}}]{Robertson2013}
{Robertson}, B.~E., {Furlanetto}, S.~R., {Schneider}, E., {et~al.} 2013, \apj,
  768, 71

\bibitem[{{Salpeter}(1955)}]{Salpeter1955}
{Salpeter}, E.~E. 1955, \apj, 121, 161

\bibitem[{{Schaerer} {et~al.}(2019){Schaerer}, {Fragos}, \&
  {Izotov}}]{Schaerer2019}
{Schaerer}, D., {Fragos}, T., \& {Izotov}, Y.~I. 2019, \aap, 622, L10

\bibitem[{{Schreiber} {et~al.}(2018{\natexlab{a}}){Schreiber}, {Labb{\'e}},
  {Glazebrook}, {Bekiaris}, {Papovich}, {Costa}, {Elbaz}, {Kacprzak},
  {Nanayakkara}, \& {Oesch}}]{Schreiber2018b}
{Schreiber}, C., {Labb{\'e}}, I., {Glazebrook}, K., {et~al.}
  2018{\natexlab{a}}, \aap, 611, A22

\bibitem[{{Schreiber} {et~al.}(2018{\natexlab{b}}){Schreiber}, {Glazebrook},
  {Nanayakkara}, {Kacprzak}, {Labb{\'e}}, {Oesch}, {Yuan}, {Tran}, {Papovich},
  {Spitler}, \& {Straatman}}]{Schreiber2018}
{Schreiber}, C., {Glazebrook}, K., {Nanayakkara}, T., {et~al.}
  2018{\natexlab{b}}, \aap, 618, A85

\bibitem[{{Scoville} {et~al.}(2007){Scoville}, {Aussel}, {Brusa}, {Capak},
  {Carollo}, {Elvis}, {Giavalisco}, {Guzzo}, {Hasinger}, {Impey}, {Kneib},
  {LeFevre}, {Lilly}, {Mobasher}, {Renzini}, {Rich}, {Sanders}, {Schinnerer},
  {Schminovich}, {Shopbell}, {Taniguchi}, \& {Tyson}}]{Scoville2007}
{Scoville}, N., {Aussel}, H., {Brusa}, M., {et~al.} 2007, \apjs, 172, 1

\bibitem[{{Senchyna} {et~al.}(2017){Senchyna}, {Stark}, {Vidal-Garc{\'\i}a},
  {Chevallard}, {Charlot}, {Mainali}, {Jones}, {Wofford}, {Feltre}, \&
  {Gutkin}}]{Senchyna2017}
{Senchyna}, P., {Stark}, D.~P., {Vidal-Garc{\'\i}a}, A., {et~al.} 2017, \mnras,
  472, 2608

\bibitem[{{Shim} {et~al.}(2011){Shim}, {Chary}, {Dickinson}, {Lin}, {Spinrad},
  {Stern}, \& {Yan}}]{Shim2011}
{Shim}, H., {Chary}, R.-R., {Dickinson}, M., {et~al.} 2011, \apj, 738, 69

\bibitem[{{Shin} {et~al.}(2008){Shin}, {Trac}, \& {Cen}}]{Shin2008}
{Shin}, M.-S., {Trac}, H., \& {Cen}, R. 2008, \apj, 681, 756

\bibitem[{{Shivaei} {et~al.}(2018){Shivaei}, {Reddy}, {Siana}, {Shapley},
  {Kriek}, {Mobasher}, {Freeman}, {Sanders}, {Coil}, {Price}, {Fetherolf},
  {Azadi}, {Leung}, \& {Zick}}]{Shivaei2018}
{Shivaei}, I., {Reddy}, N.~A., {Siana}, B., {et~al.} 2018, \apj, 855, 42

\bibitem[{{Sparre} {et~al.}(2017){Sparre}, {Hayward}, {Feldmann},
  {Faucher-Gigu{\`e}re}, {Muratov}, {Kere{\v{s}}}, \& {Hopkins}}]{Sparre2017}
{Sparre}, M., {Hayward}, C.~C., {Feldmann}, R., {et~al.} 2017, \mnras, 466, 88

\bibitem[{Spearman(1904)}]{Spearman1904}
Spearman, C. 1904, The American Journal of Psychology, 15, 72

\bibitem[{{Spitler} {et~al.}(2012){Spitler}, {Labb{\'e}}, {Glazebrook},
  {Persson}, {Monson}, {Papovich}, {Tran}, {Poole}, {Quadri}, {van Dokkum},
  {Kelson}, {Kacprzak}, {McCarthy}, {Murphy}, {Straatman}, \&
  {Tilvi}}]{Spitler2012}
{Spitler}, L.~R., {Labb{\'e}}, I., {Glazebrook}, K., {et~al.} 2012, \apjl, 748,
  L21

\bibitem[{{Spitler} {et~al.}(2014){Spitler}, {Straatman}, {Labb{\'e}},
  {Glazebrook}, {Tran}, {Kacprzak}, {Quadri}, {Papovich}, {Persson}, {van
  Dokkum}, {Allen}, {Kawinwanichakij}, {Kelson}, {McCarthy}, {Mehrtens},
  {Monson}, {Nanayakkara}, {Rees}, {Tilvi}, \& {Tomczak}}]{Spitler2014}
{Spitler}, L.~R., {Straatman}, C.~M.~S., {Labb{\'e}}, I., {et~al.} 2014, \apjl,
  787, L36

\bibitem[{{Stark} {et~al.}(2015){Stark}, {Walth}, {Charlot}, {Cl{\'e}ment},
  {Feltre}, {Gutkin}, {Richard}, {Mainali}, {Robertson}, {Siana}, {Tang}, \&
  {Schenker}}]{Stark2015c}
{Stark}, D.~P., {Walth}, G., {Charlot}, S., {et~al.} 2015, \mnras, 454, 1393

\bibitem[{{Stark} {et~al.}(2017){Stark}, {Ellis}, {Charlot}, {Chevallard},
  {Tang}, {Belli}, {Zitrin}, {Mainali}, {Gutkin}, {Vidal-Garc{\'{\i}}a},
  {Bouwens}, \& {Oesch}}]{Stark2017}
{Stark}, D.~P., {Ellis}, R.~S., {Charlot}, S., {et~al.} 2017, \mnras, 464, 469

\bibitem[{{Steidel} {et~al.}(2018){Steidel}, {Bogosavlevic}, {Shapley},
  {Reddy}, {Rudie}, {Pettini}, {Trainor}, \& {Strom}}]{Steidel2018}
{Steidel}, C.~C., {Bogosavlevic}, M., {Shapley}, A.~E., {et~al.} 2018, ArXiv
  e-prints, arXiv:1805.06071

\bibitem[{{Steidel} {et~al.}(2014){Steidel}, {Rudie}, {Strom}, {Pettini},
  {Reddy}, {Shapley}, {Trainor}, {Erb}, {Turner}, {Konidaris}, {Kulas}, {Mace},
  {Matthews}, \& {McLean}}]{Steidel2014}
{Steidel}, C.~C., {Rudie}, G.~C., {Strom}, A.~L., {et~al.} 2014, \apj, 795, 165

\bibitem[{{Straatman} {et~al.}(2014){Straatman}, {Labb{\'e}}, {Spitler},
  {Allen}, {Altieri}, {Brammer}, {Dickinson}, {van Dokkum}, {Inami},
  {Glazebrook}, {Kacprzak}, {Kawinwanichakij}, {Kelson}, {McCarthy},
  {Mehrtens}, {Monson}, {Murphy}, {Papovich}, {Persson}, {Quadri}, {Rees},
  {Tomczak}, {Tran}, \& {Tilvi}}]{Straatman2014}
{Straatman}, C.~M.~S., {Labb{\'e}}, I., {Spitler}, L.~R., {et~al.} 2014, \apjl,
  783, L14

\bibitem[{{Straatman} {et~al.}(2016){Straatman}, {Spitler}, {Quadri},
  {Labb{\'e}}, {Glazebrook}, {Persson}, {Papovich}, {Tran}, {Brammer},
  {Cowley}, {Tomczak}, {Nanayakkara}, {Alcorn}, {Allen}, {Broussard}, {van
  Dokkum}, {Forrest}, {van Houdt}, {Kacprzak}, {Kawinwanichakij}, {Kelson},
  {Lee}, {McCarthy}, {Mehrtens}, {Monson}, {Murphy}, {Rees}, {Tilvi}, \&
  {Whitaker}}]{Straatman2016}
{Straatman}, C. M.~S., {Spitler}, L.~R., {Quadri}, R.~F., {et~al.} 2016, \apj,
  830, 51

\bibitem[{{Strom} {et~al.}(2017){Strom}, {Steidel}, {Rudie}, {Trainor},
  {Pettini}, \& {Reddy}}]{Strom2017}
{Strom}, A.~L., {Steidel}, C.~C., {Rudie}, G.~C., {et~al.} 2017, \apj, 836, 164

\bibitem[{{Szokoly} {et~al.}(2004){Szokoly}, {Bergeron}, {Hasinger}, {Lehmann},
  {Kewley}, {Mainieri}, {Nonino}, {Rosati}, {Giacconi}, {Gilli}, {Gilmozzi},
  {Norman}, {Romaniello}, {Schreier}, {Tozzi}, {Wang}, {Zheng}, \&
  {Zirm}}]{Szokoly2004}
{Szokoly}, G.~P., {Bergeron}, J., {Hasinger}, G., {et~al.} 2004, \apjs, 155,
  271

\bibitem[{{Tang} {et~al.}(2018){Tang}, {Stark}, {Chevallard}, \&
  {Charlot}}]{Tang2018}
{Tang}, M., {Stark}, D., {Chevallard}, J., \& {Charlot}, S. 2018, arXiv
  e-prints, arXiv:1809.09637

\bibitem[{{Tomczak} {et~al.}(2016){Tomczak}, {Quadri}, {Tran}, {Labb{\'e}},
  {Straatman}, {Papovich}, {Glazebrook}, {Allen}, {Brammer}, {Cowley},
  {Dickinson}, {Elbaz}, {Inami}, {Kacprzak}, {Morrison}, {Nanayakkara},
  {Persson}, {Rees}, {Salmon}, {Schreiber}, {Spitler}, \&
  {Whitaker}}]{Tomczak2016}
{Tomczak}, A.~R., {Quadri}, R.~F., {Tran}, K.-V.~H., {et~al.} 2016, \apj, 817,
  118

\bibitem[{{Tran} {et~al.}(2015){Tran}, {Nanayakkara}, {Yuan}, {Kacprzak},
  {Glazebrook}, {Kewley}, {Momcheva}, {Papovich}, {Quadri}, {Rudnick},
  {Saintonge}, {Spitler}, {Straatman}, \& {Tomczak}}]{Tran2015}
{Tran}, K.-V.~H., {Nanayakkara}, T., {Yuan}, T., {et~al.} 2015, \apj, 811, 28

\bibitem[{{Urrutia} {et~al.}(2019){Urrutia}, {Wisotzki}, {Kerutt}, {Schmidt},
  {Herenz}, {Klar}, {Saust}, {Werhahn}, {Diener}, {Caruana}, {Krajnovi{\'c}},
  {Bacon}, {Boogaard}, {Brinchmann}, {Enke}, {Maseda}, {Nanayakkara},
  {Richard}, {Steinmetz}, \& {Weilbacher}}]{Urrutia2019}
{Urrutia}, T., {Wisotzki}, L., {Kerutt}, J., {et~al.} 2019, \aap, 624, A141

\bibitem[{{Wang} {et~al.}(2019){Wang}, {Schreiber}, {Elbaz}, {Yoshimura},
  {Kohno}, {Shu}, {Yamaguchi}, {Pannella}, {Franco}, {Huang}, {Lim}, \&
  {Wang}}]{Wang2019}
{Wang}, T., {Schreiber}, C., {Elbaz}, C., {et~al.} 2019, arXiv e-prints,
  arXiv:1908.02372

\bibitem[{{Wilkins} {et~al.}(2016){Wilkins}, {Feng}, {Di-Matteo}, {Croft},
  {Stanway}, {Bouwens}, \& {Thomas}}]{Wilkins2016b}
{Wilkins}, S.~M., {Feng}, Y., {Di-Matteo}, T., {et~al.} 2016, \mnras, 458, L6

\bibitem[{{Williams} {et~al.}(2009){Williams}, {Quadri}, {Franx}, {van Dokkum},
  \& {Labb{\'e}}}]{Williams2009}
{Williams}, R.~J., {Quadri}, R.~F., {Franx}, M., {van Dokkum}, P., \&
  {Labb{\'e}}, I. 2009, \apj, 691, 1879

\bibitem[{{Yuan} {et~al.}(2014){Yuan}, {Nanayakkara}, {Kacprzak}, {Tran},
  {Glazebrook}, {Kewley}, {Spitler}, {Poole}, {Labb{\'e}}, {Straatman}, \&
  {Tomczak}}]{Yuan2014}
{Yuan}, T., {Nanayakkara}, T., {Kacprzak}, G.~G., {et~al.} 2014, \apjl, 795,
  L20

\end{thebibliography}

\end{document}